%% file: template.tex
\documentclass{article}

\usepackage{arxiv}

\usepackage[utf8]{inputenc} 
\usepackage[T1]{fontenc}    
\usepackage{hyperref}       
\usepackage{url}            
\usepackage{booktabs}       
\usepackage{amsfonts}       
\usepackage{nicefrac}       
\usepackage{microtype}      
\usepackage{graphicx}
\usepackage{natbib}
\usepackage{doi}
\usepackage{amsmath}
\usepackage{amssymb}   
\usepackage{dsfont}

\title{Prior-informed conditional Gaussian graphical models: an application to protein interaction network reconstruction}

\author{
Alessia Mapelli$^{1,2}$,
Michela Carlotta Massi$^{2}$,
Gianmauro Cuccuru$^{2}$, Emanuele Di Angelantonio$^{2,3,4,5,6}$,
Francesca Ieva$^{1,2}$
}

\date{
$^{1}$ MOX, Department of Mathematics, Politecnico di Milano, Milan, Italy\\
$^{2}$ Health Data Science Research Centre, Human Technopole, Milan, Italy\\
$^{3}$ British Heart Foundation Cardiovascular Epidemiology Unit, Department of Public Health and Primary Care, University of Cambridge, Cambridge, UK\\
$^{4}$ Victor Phillip Dahdaleh Heart and Lung Research Institute, University of Cambridge, Cambridge, UK\\
$^{5}$ British Heart Foundation Centre of Research Excellence, University of Cambridge, Cambridge, UK\\
$^{6}$ National Institute for Health and Care Research Blood and Transplant Research Unit in Donor Health and Behaviour, University of Cambridge, Cambridge, UK\\[6pt]
\texttt{alessia.mapelli@polimi.it}
}

\renewcommand{\shorttitle}{Prior-informed conditional GGMs}

\hypersetup{
pdftitle={Prior-informed conditional Gaussian graphical models: an application to protein interaction network reconstruction},
pdfsubject={q-bio.NC, q-bio.QM},
pdfauthor={Alessia Mapelli, Michela Carlotta Massi, Gianmauro Cuccuru,Emanuele Di Angelantonio, Francesca Ieva},
pdfkeywords={Gaussian graphical models, prior knowledge integration,
conditional graphical models, protein-protein interaction networks, network-based biomarker discovery},
}

\begin{document}
\maketitle

\begin{abstract}
	\textbf{Motivation:} Protein-protein interaction (PPI) networks, estimated
from high-throughput omics data, foster biomarker discovery and precision medicine. Gaussian graphical models (GGMs) offer a principled reconstruction framework. Yet, existing applications face two limitations: they overlook the rich existing knowledge encoded in curated biological databases, and they assume a homogeneous network structure across all individuals, neglecting the influence of covariates or confounding factors on these interactions and preventing personalised representations. Even though these limitations have been addressed separately in previous work, no current approach resolves them simultaneously.\\
\textbf{Results:} We introduce a prior-informed conditional Gaussian graphical model that integrates database-derived interaction priors with covariate-dependent network modeling in a unified, scalable framework. The key methodological innovation is a structured, weighted penalty that selectively incorporates priors into population-level network estimation, while leaving context-specific perturbations entirely data-driven, as curated databases capture canonical interactions rather than disease-specific signals. Simulation studies demonstrate consistent and
robust improvements in population-level network reconstruction across
diverse settings, even when prior knowledge is imperfect. Applied to UK Biobank cardiometabolic proteomics ($n\!=\!49{,}129$, $p\!=\!366$ proteins), the method recovers T2D-associated network perturbations, identifying 34 network-central candidate biomarkers, several detectable only through their connectivity, not differential expression, and revealing six biologically coherent protein communities with distinct pathway enrichments spanning metabolic, cardiovascular, and cancer-related processes.\\
\textbf{Availability and implementation:} Code is available at
\url{https://github.com/AlessiaMapelli/Prior-informed-conditional-GGMs}.
\end{abstract}


\section{Introduction}
\input{Intro}

\section{Methods}
\label{sec:methods}
\input{Methods}

\section{Experimental Setup}
\input{ExperimentalSetup}

\section{Results}
\input{Results}

\section{Discussion and Conclusion}
\input{Discussion}


\section*{Author contributions}

A.M.\ conceived and designed the methodology, developed the software,
conducted all simulation analyses and the case study, and wrote the
manuscript. M.C.M.\ contributed to data curation, biological interpretation of results, and critically revised the manuscript. G.C.\ contributed to computational
infrastructure and data access. E.D.A.\ supervised the study, providing clinical expertise. F.I.\ supervised the study and contributed to methodology design.
All authors reviewed and approved the final version.

\section*{Supplementary data}

Supplementary data are available at the bottom of this file, including: methodological details of the estimation pipeline (Section~S1); additional simulation results across all 24 configurations (Section~S2);
full data partitioning and preprocessing details in T2D case study (Section~S3); formal definitions of centrality measures (Section~S4);
T2D risk prediction model details (Section~S5); per-configuration numerical summaries across all 24 simulation configurations (Table~S1); summary of the characteristics of the UK Biobank  cohort and partitioning (Table~S2);
complete list of all 34 network-central proteins with centrality scores
(Table~S3); community membership assignments and full enrichment results
(Table~S4); ROC curves for the T2D risk prediction model (Figure~S1); subnetwork of top-ranked central proteins (Figure~S2); Community structure of the T2D-associated protein interaction network (Figure~S3).

\section*{Conflict of interest}

None declared.

\section*{Acknowledgements}

The present research has been supported by MUR, grant Dipartimento di
Eccellenza 2023--2027. F.\ Ieva acknowledges the National Plan for NRRP
Complementary Investments ``Advanced Technologies for Human-centred
Medicine'' (PNC0000003). This study used data from the UK Biobank Resource under Application Number 102297.
Large language models were used as an aid for code writing and to correct written text.

\section*{Data availability}

Code and computational templates are available at
\url{https://github.com/AlessiaMapelli/Prior-informed-conditional-GGMs}.


\bibliographystyle{unsrtnat}
\bibliography{references}  

\input{Supplementary}

\end{document}

%% file: Intro.tex
High-throughput experimental technologies now generate large-scale omics measurements across thousands of individuals, creating unprecedented opportunities for systems-level disease modelling through biological network reconstruction. These networks, where nodes represent molecular entities and edges their associations, provide a natural framework for studying the functional organisation of biological systems and have become essential tools for biomarker discovery, disease stratification, and therapeutic target identification~\citep{pavlopoulos2011using,choobdar2019assessment}. This systems-level perspective requires analytical approaches capable of modeling complex molecular interaction systems and their disease-driven perturbations. Within this framework, protein-protein interaction (PPI) networks represent a particularly compelling application. Proteins occupy a critical middle layer between the genome and the environment, constitute the dominant target class in omics-based drug discovery, enabling clinical translation, and benefit from large-scale curated databases such as STRING~\citep{szklarczyk2021string}, which aggregate decades of experimental evidence on interaction confidence.

Gaussian graphical models (GGMs) have become one of the predominant frameworks for data-driven network reconstruction: under the multivariate normality assumption, the precision matrix $\boldsymbol{\Omega} = \boldsymbol{\Sigma}^{-1}$ directly encodes conditional independence between variables, and the network edge set is defined as $E = \{(j,k): \boldsymbol{\Omega}_{j,k} \neq 0\}$~\citep{shutta2022gaussian}. Neighbourhood selection~\citep{meinshausen2006variable} methods decompose the precision matrix estimation into $p$ parallel Lasso regressions, enabling scalable and node-specific regularisation, a crucial advantage for biological networks whose scale-free topology produces heterogeneous local degree distributions~\citep{koutrouli2020guide}. Despite their popularity, standard GGM applications exhibit two important limitations. First, they treat all potential edges equally during regularisation, ignoring the wealth of prior knowledge encoded in curated databases. Incorporating such knowledge via weighted penalties has been shown to substantially improve reconstruction accuracy~\citep{wang2013incorporating}. Second, conventional GGMs assume a homogeneous network structure across samples, neglecting individual-level variability. In many biological and disease contexts, heterogeneity is a defining feature. Subject-level covariates, such as age, genotype, or clinical characteristics, may modulate not only expression levels but also the structure of molecular interactions~\citep{xie2022conditional}. Conditional GGM frameworks have recently addressed this second limitation by allowing the mean and the precision matrix to depend linearly on external covariates through interaction-based neighborhood selection methods~\citep{zhang2023high}. These two methodological advances, however, have largely evolved independently. To our knowledge, no current framework simultaneously integrates database-derived prior knowledge and covariate-dependent network estimation within a unified and scalable inference procedure. Existing implementations are also often limited in accessibility and computational flexibility, restricting their broad applicability.

We address both limitations by introducing a prior-informed conditional Gaussian graphical model. Building on the conditional neighborhood selection framework of~\citep{zhang2023high}, we extend node-wise penalized regression with a structured weighted penalty that distinguishes between population-level and covariate-specific network components. Database-derived prior weights are included in the population-level estimation, as curated databases reflect canonical interactions under standard conditions. In contrast, covariate-specific perturbations are estimated exclusively from data. Our approach thus simultaneously leverages accumulated biological knowledge to achieve robust population-level reconstruction while maintaining sensitivity to novel disease- or context-specific signals that may not be represented in existing databases.
In addition to the methodological contribution, we provide an accessible and scalable R implementation that automatically adapts from standard neighbourhood-based network reconstruction to the complete prior-informed conditional framework depending on the available inputs, and supports execution from personal computers to HPC environments.

%% file: Methods.tex
\subsection{Gaussian graphical models and neighborhood selection}
\label{subsec:ggm}

Graphical models represent statistical dependencies among variables using a graph $\mathcal{G} = (V, E)$, where $V = \{1, \dots, p\}$ is the set of nodes corresponding to $p$ variables, and $E$ is the set of edges denoting pairwise relationships. Let $Y = (Y_1, \ldots, Y_p)^T \sim \mathcal{N}_p(\mu,\boldsymbol{\Sigma})$ represent $p$ molecular expression levels. In a GGM, the precision matrix $\boldsymbol{\Omega} = \boldsymbol{\Sigma}^{-1}$ encodes conditional independence between the variables: $\boldsymbol{\Omega}_{j,k} = 0$ if and only if $Y_j \perp Y_k \mid Y_{-(j,k)}$, and the network edges are defined as $E = \{(j,k): \boldsymbol{\Omega}_{j,k} \neq 0\}$~\citep{shutta2022gaussian} with edge weights linked to partial correlations
\begin{equation}
    \rho_{j,k|-j,k} = - \frac{\Omega_{j,k}}{\sqrt{\Omega_{j,j}\Omega_{k,k}}}. \label{eq1}
\end{equation}
The sparsity pattern of $\Omega$ thus defines the network structure, and estimating the GGM is equivalent to estimating $\Omega$.
The precision matrix $\Omega$ can be effectively estimated using a neighborhood selection approach~\citep{meinshausen2006variable}, which decomposes precision matrix estimation into $p$ independent Lasso regressions. Without loss of generality and for ease of notation in the following, we set $j=p$:
\begin{equation}
\hat{\delta}_p = \arg\min_{\delta}\;
\frac{1}{2n}\bigl\| Y_p - Y_{-p}\delta \bigr\|_2^2 + \lambda\|\delta\|_1,\label{eq:neighsel}
\end{equation}
where $Y_{-p}$ contains all variables except $Y_p$ and regression
coefficients relate to the precision matrix as
$\hat{\delta}_{p,k} = -\Omega_{p,k}/\Omega_{p,p}$ so that the sparsity pattern of the network can be directly retrieved from the estimated regression coefficients. Partial correlations and weights can be recovered from symmetrisation procedures for node-pair estimates~\citep{wang2013incorporating}. This parallelisable decomposition is computationally advantageous and enables node-specific sparsity control, which is crucial for biological networks with heterogeneous scale-free topologies~\citep{koutrouli2020guide}.

\subsection{Incorporating prior biological knowledge}
\label{subsec:prior}

Let $\mathbf{W} \in [0,1]^{p \times p}$ be a symmetric weight matrix where $W_{j,k}$ encodes prior confidence that entities $j$ and $k$ interact (e.g.\ the STRING combined confidence score for proteins). Under the Bayesian interpretation of Lasso, where the $\ell_1$ penalty corresponds to a Laplacian prior on regression coefficients, differential penalties across edges correspond to mixture priors with edge-specific shrinkage~\citep{wang2013incorporating}. Weighted neighborhood selection incorporates this by replacing the uniform penalty in Eq.~\ref{eq:neighsel} with:
\begin{equation}
\hat{\delta}_p = \arg\min_{\delta}\;
  \frac{1}{2n}\bigl\| Y_p - Y_{-p}\delta \bigr\|_2^2 + \lambda\bigl\|(1-\mathbf{W}_{p,-p})\odot\delta\bigr\|_1,
\label{eq:neighsel_prior}
\end{equation}
applying weaker penalties to edges with strong prior support
($W_{p,k}\approx 1$) and stronger shrinkage to the unsupported
pairs~\citep{zuo2017incorporating}.

\subsection{Conditional Gaussian graphical models}
\label{subsec:cgm}

To model individual-level heterogeneity, let $X$ denote a $q$-dimensional vector of external covariates (e.g.\ disease status, sex, age). Following~\citep{zhang2023high}, we assume $Y \mid X \sim \mathcal{N}_p(\mathbf{B}X,\,\boldsymbol{\Omega}(X)^{-1})$ where both the mean and the precision matrix depend linearly on the external covariates:
\begin{equation}
\boldsymbol{\Omega}(X) = \boldsymbol{\Delta}^0 + \sum_{h=1}^{q} X_h\boldsymbol{\Delta}^h.
\label{eq:conditional}
\end{equation}
In the previous equation, $\boldsymbol{\Delta}^0 \in \mathrm{Sym}(p)$ is the baseline
population-level network and $\boldsymbol{\Delta}^h \in \mathrm{Sym}(p)$ encodes how covariate $X_h$ modulates network structure. The model can, once more, be decomposed into a sequence of regression problems that include the interactions between $Y$ and external covariates $X$:
\begin{equation}
Y_p = X^T \beta_p + \sum_{k \neq p} \delta^0_{pk}(Y_k - X^T \beta_k) + \sum_{k \neq p} \sum_{h=1}^q \delta^h_{pk} X_h (Y_k - X^T \beta_k) + \varepsilon_p,
\label{eq:neigh_cond_eq}
\end{equation}
where $\beta_p$ captures covariate effects on the mean levels, while $\delta^0_{pk}$ and $\delta^h_{pk}$ represent, respectively, the population-level presence and the covariate-specific modulation of the edge $(p,k)$. The error term follows $\varepsilon_p \sim \mathcal{N}(0, \sigma_{pp}^{-1})$. This formulation allows both the presence and strength of network edges to vary across individuals, enabling personalised network reconstruction.

\subsection{Prior-informed conditional graphical Lasso}
\label{subsec:method}

We extend conditional neighbourhood selection to incorporate prior
knowledge through a structured weighted penalty. We first estimate covariate effects on the mean levels $\mu(\mathbf{X})$ via standard Lasso:
\begin{equation}
\hat{\beta}_p = \arg\min_{\beta} \frac{1}{2n} \|Y_p - X\beta\|_2^2 + \lambda_p^B \|\beta\|_1.
\label{eq:our_neig_selec_step1}
\end{equation}
After regressing out mean covariate effects, $Z_j = Y_j - X\hat{\beta}j$ for all $j$, the model in Eq.~\ref{eq:neigh_cond_eq} reduces to a set of node-wise regressions on residuals and their interactions with the covariates. We then jointly estimate the baseline coefficients, $d^0_p=(\delta^0{p,1}, \ldots, \delta^0_{p,p-1})^T \in \mathbb{R}^{p-1}$, and the covariate-specific coefficients. $d^h_p=(\delta^h_{p,1}, \ldots, \delta^h_{p,p-1})^T \in \mathbb{R}^{p-1}$ for $h = 1,\ldots,q$, by concatenating them into a single vector
$d_p=[(d^0_p)^T,(d^1_p)^T, \ldots ,(d^q_p)^T ]^T$ and solving:
\begin{equation}
\hat{d}_p = \arg\min_{d_p}\;
  \frac{1}{2n}\bigl\|Z_p - \mathbf{U}_{-p}d_p\bigr\|_2^2 + \alpha\lambda_p^\Delta
     \bigl\|(1-\mathbf{W}_{p,-p})\odot d^0_p\bigr\|_1 + \vartheta\alpha\lambda_p^\Delta \bigl\|d^{-0}_p\bigr\|_1 + (1-\alpha)\lambda_p^\Delta \sum_{h=1}^{q}\bigl\|d^h_p\bigr\|_2, \label{eq:main}
\end{equation}
where $d^{-0}_p$ collects all
covariate-specific parameters and the design matrix $\mathbf{U}_{-p} = [1, X^T]^T Z_{-p}$ collects the residuals and their interactions with the covariates (excluding node $p$). Note that, each node-wise regression involves approximately $p\times(q+1)$ coefficients, which in typical omics applications substantially exceeds the sample size ($p\times(q+1)\gg n$). Regularisation is therefore necessary to ensure identifiability and stable estimation, independently of the underlying network density.\\
The three penalty terms encode distinct structural assumptions.
(i) \textit{Prior-informed sparsity}: the weighted $\ell_1$ penalty is applied only to baseline effects $d^0_p$, reducing shrinkage on database-supported interactions. (ii) \textit{Element-wise sparsity}: a uniform $\ell_1$ penalty on covariate effects $d^{-0}_p$ promotes sparse context-dependent modulations.(iii) \textit{Group sparsity}: the $\ell_2$ group penalty encourages all
effects associated with a single covariate to be jointly zero or non-zero, enabling covariate selection.
The key design choice is that prior weights are applied exclusively to $d^0_p$ and not to $d^h_p$. This is biologically principled as
curated databases document canonical interactions under standard conditions, whereas disease- or context-specific perturbations represent condition-dependent phenomena that should be inferred directly from data. Applying prior weights to covariate effects would introduce a systematic bias that suppresses the detection of novel, context-specific interactions.
The parameter $\lambda_p^\Delta > 0$ controls the overall strength of regularisation, $\alpha\in[0,1]$ balances group and element-wise sparsity, and $\vartheta>0$ regulates the relative penalty applied to covariate-specific effects versus prior-weighted baseline edges.

As the node-wise regressions usually yield asymmetric estimates $\hat{\delta}^h_{jk}$ and $\hat{\delta}^h_{kj}$, the edge set linked to baseline effects and covariate-specific effects $E^h$ with $h = 0,\ldots,q$ is constructed via a symmetrization strategy depending on the level of desired sparsity~\citep{zhang2023high}.
The \textit{OR rule}, used throughout the main analyses, retains edges if detected in either direction:
\begin{equation}
E^h = \{(p,k): \delta^h_{p,k} \neq 0\ | \delta^h_{k,p} \neq 0\}
\label{eq:sym}
\end{equation}
Edge weights are similarly retrieved for each pair of nodes $(p,k)$: $$\hat{\rho}^h_{pk|-\{p,k\}} =  \hat{\delta}^h_{p,k}\mathds{1}_{|\hat{\delta}^h_{p,k}| \geq |\hat{\delta}^h_{k,p}|} + \hat{\delta}^h_{k,p}\mathds{1}_{|\hat{\delta}^h_{p,k}| < |\hat{\delta}^h_{k,p}|}.$$
Thanks to the linear parameterization in Eq.~\ref{eq:conditional}, the personalized adjacency matrix for an individual $i$ with
covariate vector $x_i$ is obtained by:
\begin{equation}
\hat{\mathbf{A}}(x_i)_{p,k} = \hat{\rho}^0_{pk|-\{p,k\}} + \sum_{h=1}^{q} x_{ih}\hat{\rho}^h_{pk|-\{p,k\}}.
\label{eq:personalized}
\end{equation}

Further methodological details are provided in Supplementary Section~S1.

\subsection{Computational implementation}
\label{subsec:computation}

\begin{figure*}[htbp]
\centering
\includegraphics[width=\textwidth]{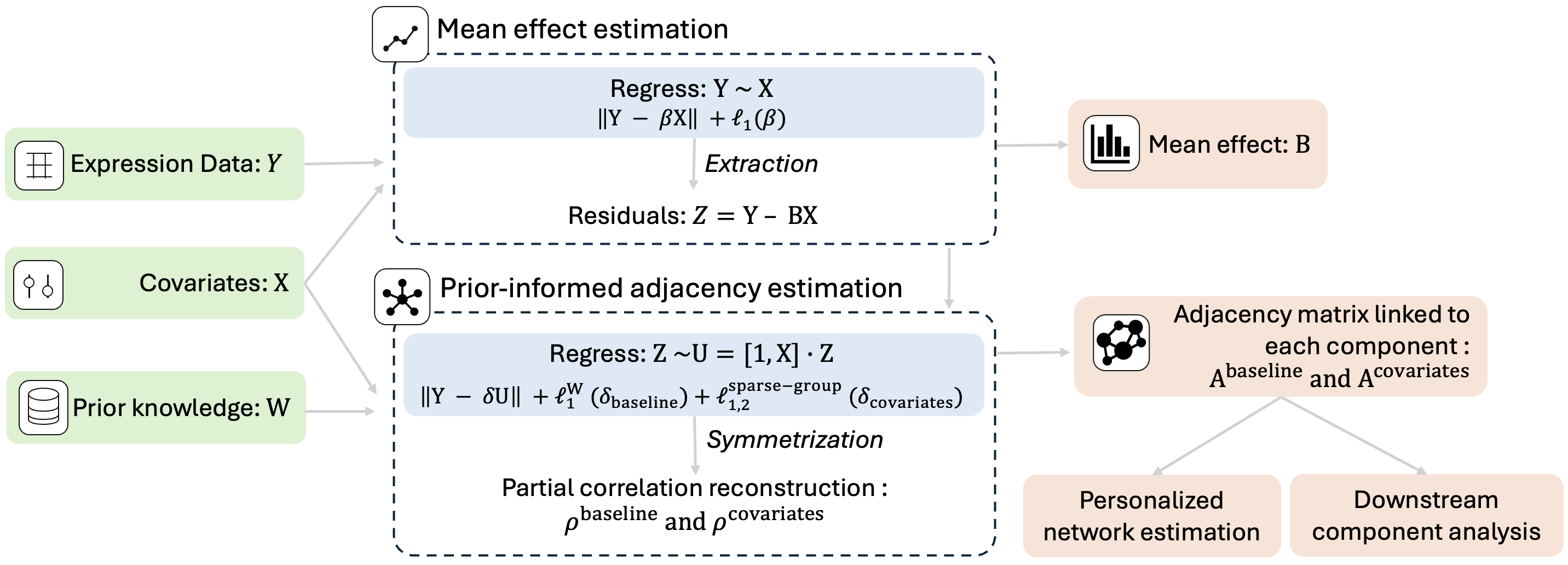}
\caption{Overview of the prior-informed conditional GGM estimation
pipeline. \textbf{(Step 1)} Node-wise Lasso regression of expression
data on covariates yields residuals $Z$, retrieving the effect of each covariate on protein mean expression.
\textbf{(Step 2)} For each node, the weighted conditional regression
(Eq.~\ref{eq:main}) jointly estimates the baseline neighborhood (the
baseline effects of each node on the response node), penalized with
prior-derived weights, and the covariate-specific neighborhood, penalized uniformly in a sparse-group fashion to preserve
data-driven discovery. Partial correlation weights for each node pair in each condition
are reconstructed through symmetrization; component-specific adjacency
matrices are constructed with elements corresponding to pairwise partial
correlations. Downstream, personalized networks are obtained as a
linear combination of the component adjacency matrices
(Eq.~\ref{eq:personalized}), and differential network analyses, such as node centrality
and community detection, can be performed component-wise.}\label{fig:pipeline}
\end{figure*}

The method described above and schematized in Figure~\ref{fig:pipeline} has been implemented in R and is available at
\url{https://github.com/AlessiaMapelli/Prior-informed-conditional-GGMs} with usage examples.\\
The pipeline is accessed via a single entry point
function \texttt{GGReg\_full\_estimation}, requiring only expression data
and, optionally, covariates and a prior weight matrix. Depending on the inputs provided, the function automatically fits the appropriate model:
from a standard neighbourhood-based network reconstruction (no covariates, no prior) through to the full
prior-informed conditional model, making the same interface useful across diverse research contexts. Regularisation parameters $\lambda_p^B$ and $\lambda_p^\Delta$ are selected via cross-validation using \texttt{glmnet}~\citep{friedman2010regularization} and
\texttt{sparsegl}~\citep{liang2024sparsegl}; structural parameters
$\alpha$ and $\vartheta$ are optimised by BIC over a predefined grid, with optional random search for large parameter spaces. To accommodate datasets ranging from pilot studies to population biobanks, the pipeline provides three execution strategies: sequential processing ($p\leq150$), shared-memory parallelisation via \texttt{foreach}/\texttt{doParallel} ($p>150$), and SLURM array jobs for HPC deployment (suggested $p>1000$), with one independent job per node. A pre-screening option can further reduce the effective dimensionality of each node-wise regression when $p \times q \gg n$. Predictors lacking prior support may be excluded before estimation, either via a correlation-based criterion that removes weakly correlated node pairs following the sure independence screening principle~\citep{fan2008sure}, or via a user-defined screening matrix that enables domain-driven feature pre-selection. A companion \texttt{predict\_personalized\_network} function computes personalised adjacency matrices for new observations given their covariate vector.

%% file: ExperimentalSetup.tex
\subsection{Simulation design}
\label{subsec:sim_setup}

To systematically evaluate the proposed method and demonstrate the value of integrating domain knowledge within the conditional network estimation framework proposed in~\citep{zhang2023high}, we conducted a comprehensive simulation study across diverse scenarios. We employed a factorial design combining sample sizes $n\in\{500,1000\}$, network dimensions $p\in\{50,100\}$, covariate complexity $q\in\{1,3\}$, and three prior knowledge conditions:
(i) \textit{perfect}: complete accurate knowledge $W_{j,k} = \mathds{1}_{(j,k) \in E}$; (ii) \textit{noisy}: 80\% sensitivity, 90\% specificity with continuous confidence scores $W_{j,k}\sim\mathrm{Uniform}[0.7,1]$ for true edges and $W_{j,k}\sim[0.1,0.6]$ for false edges; (iii) \textit{no prior}: $W_{j,k} = 0$ $\forall j,k$, reflecting the model of~\citep{zhang2023high}.
Synthetic networks were generated to reflect realistic biological system topologies. Baseline precision matrices $\boldsymbol{\Delta}_0$ followed the scale-free Barab\'{a}si--Albert model~\citep{barabasi1999emergence} with power law exponent 2.5, capturing the hub-dominated connectivity patterns commonly observed in biological networks. Covariate-specific modulations $\boldsymbol{\Delta}_h$ with $h=1,...,q$ were generated using Erd\H{o}s--R\'{e}nyi random graphs~\citep{erdHos1963random} with edge probability 0.05, reflecting the expectation that external factors modulate only sparse subsets of baseline interactions.
Edge coefficients were assigned magnitudes uniformly sampled from $[0.35, 0.5]$ and random signs, ensuring sufficient signal strength while maintaining the invertibility of the precision matrices.
For each scenario, we generated subject-level data following the conditional Gaussian model $Y \mid X \sim \mathcal{N}_p(0, \boldsymbol{\Sigma}(X))$ where precision matrices varied linearly with covariates: $\boldsymbol{\Omega}(X) = \boldsymbol{\Delta}_0 + \sum_{h=1}^q X_h \boldsymbol{\Delta}_h$. Covariate vectors combined continuous variables (standardized normal) and binary indicators (Bernoulli with probability 0.5), representing realistic mixed-type external factors.
Each of the 24 configurations was replicated 10 times. Performance was assessed via accuracy, F1-score, and magnitude preservation (Spearman correlation between estimated and true precision matrix entries), evaluated separately for baseline ($\boldsymbol{\Delta}^0$) and covariate-specific ($\boldsymbol{\Delta}^h$) components. Structural hyperparameters were optimised via random search over $\alpha\in\{0.5,0.75,0.9,0.99\}$ and $\vartheta\in\{0.8,1.0,1.1,1.3\}$ using BIC.

\subsection{UK Biobank cardiometabolic proteomics}
\label{subsec:ukb_setup}

We applied the framework to proteomics data from the UK Biobank Plasma Proteome Project~\citep{sun2023plasma}, focusing on the Olink Cardiometabolic I and II panels across 49,129 participants. These panels profile proteins selected for their relevance to cardiometabolic processes spanning metabolic, cardiovascular, immune, and inflammatory pathways. Crucially, while a subset of proteins have established associations with T2D, the majority are associated with related cardiometabolic conditions without direct T2D annotation. This makes the combined panel an ideal test case: a biologically motivated candidate pool where some biomarkers are known and can serve as positive controls, while others represent genuinely unexplored candidates. After quality control (exclusion of proteins with $>10\%$ missing rate and mean imputation), 366 proteins were retained.

The objective of this case study is to demonstrate the translational utility of the method in a classic biomarker discovery task: identify T2D-associated candidate proteins and functionally coherent pathways by leveraging the prior-informed conditional network structure.
Incident type 2 diabetes (T2D) within 5 years was ascertained through ICD-10 codes, HbA1c measurements $>48\,\mathrm{mmol/mol}$, or antidiabetic drug prescriptions. Prior knowledge was derived from STRING v12.0~\citep{szklarczyk2021string}, providing a normalised $366\!\times\!366$ weight matrix from combined interaction evidence. Sex and T2D incidence were included as covariates, with sex selected based on explaining $\approx$50\% of proteomic variance in the principal component analysis. The network was estimated on 80\% of the cohort ($n=39{,}304$); a balanced subset ($n=270$) with a case-control design was used for biomarker-based risk model training, and an independent cohort ($n=9{,}555$) for testing. Full data partitioning and preprocessing details are provided in Supplementary Section~S3 and Table~S2.

%% file: Results.tex
\subsection{Simulation results}
\label{subsec:sim_res}

\begin{figure}[htbp]
\centering
\includegraphics[width=0.8\textwidth]{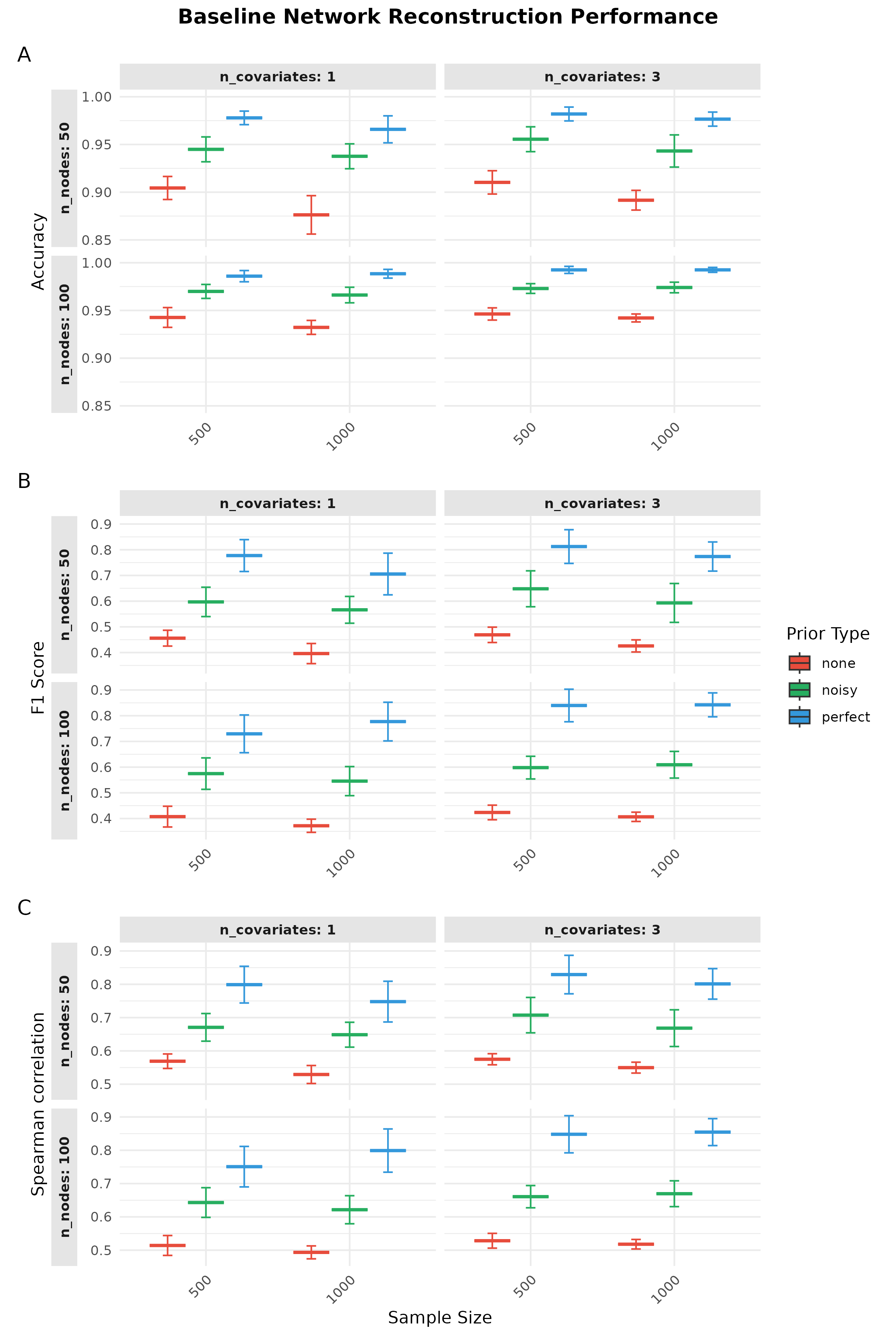}
\caption{Baseline network reconstruction performance across prior
knowledge conditions. \textbf{(A)} Accuracy, \textbf{(B)} F1-score, and
\textbf{(C)} Spearman correlation (magnitude preservation) between estimated and true precision
matrix entries for three prior conditions: no
prior (red), noisy prior (green), and perfect prior (blue). Results span all combinations of $n\in\{500,1000\}$, $p\in\{50,100\}$, and $q\in\{1,3\}$. Error bars represent standard deviations over 10 replicates.}
\label{fig:sim_baseline}
\end{figure}

\begin{figure}[htbp]
\centering
\includegraphics[width=0.8\textwidth]{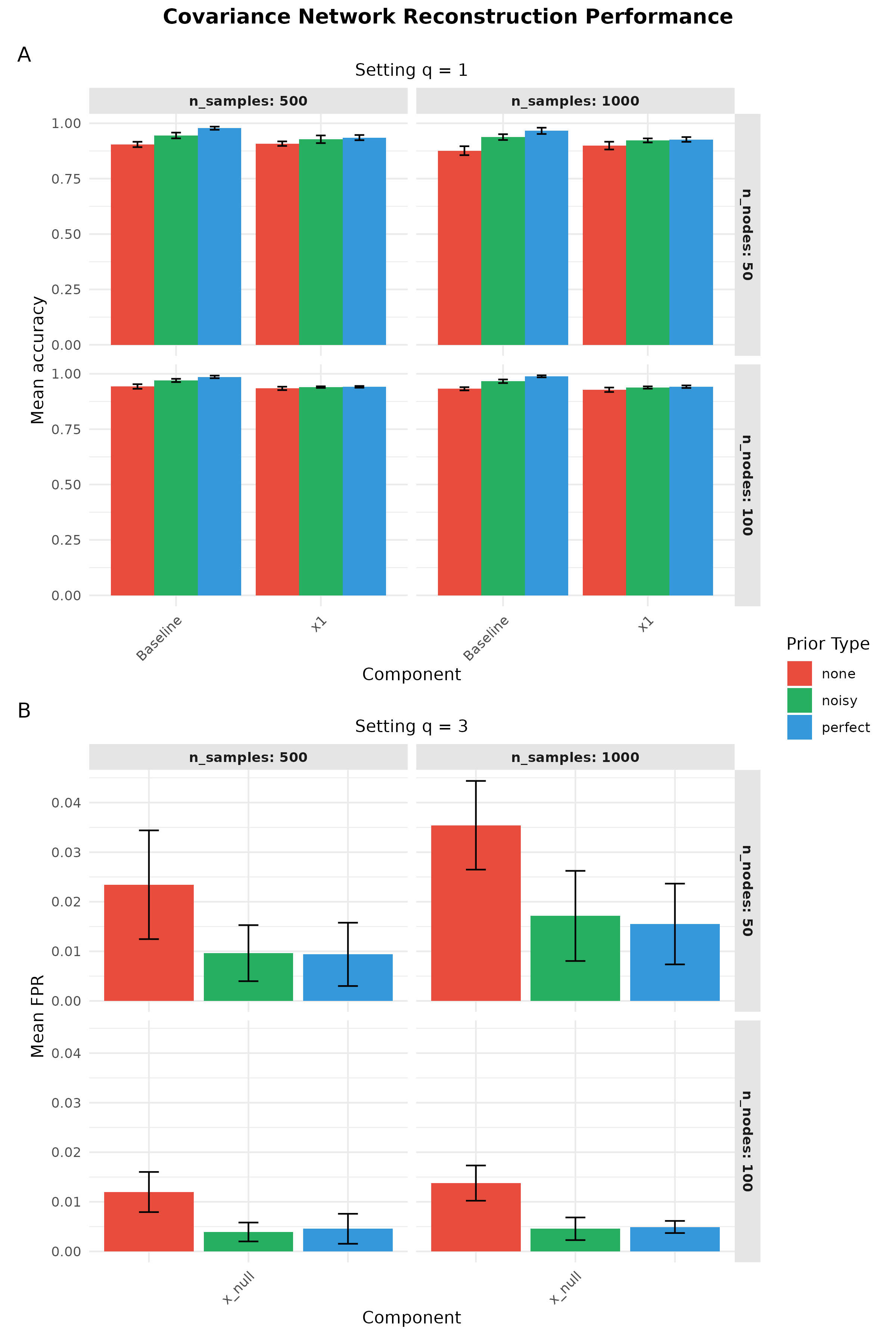}
\caption{Covariate-specific network estimation performance.
\textbf{(A)} Accuracy for baseline and covariate network components in the single-covariate scenario ($q=1$) across prior conditions and experimental settings. \textbf{(B)} False-positive rates for null covariate effects in the multi-covariate scenario ($q=3$). Error bars represent standard deviations over 10 replicates.}
\label{fig:sim_covariate}
\end{figure}

Simulation analysis results are reported in Figure~\ref{fig:sim_baseline} and Figure~\ref{fig:sim_covariate}. Incorporating prior knowledge substantially improved baseline network reconstruction across all evaluated metrics. The most pronounced gain was observed for the $F_1$-score, which increased by approximately 40\% under perfect prior knowledge, indicating a markedly improved balance between sensitivity and precision in edge recovery. Accuracy also improved consistently by approximately 5--10\%. Importantly, significant gains were retained even under the noisy prior setting, despite introducing 20\% false negatives and 10\% false positives into the prior information, demonstrating strong robustness to the realistic imperfections that characterise available biological databases
(Figure~\ref{fig:sim_baseline}A-B). The improvement in magnitude preservation was similarly notable: prior-informed methods achieved Spearman correlations of 0.7-0.9 versus 0.5-0.6 without prior information
(Figure~\ref{fig:sim_baseline}C). This enhanced quantitative accuracy in edge magnitude estimation is critical for downstream analyses such as pathway weighting and centrality-based biomarker prioritisation, where edge strength conveys biological meaning beyond binary connectivity. Performance improvements were consistent across all tested combinations of $n$, $p$, and $q$, confirming robustness to varying study designs and network dimensionality.

Reconstruction of covariate-specific network components remained stable regardless of the introduction of prior knowledge (Figure~\ref{fig:sim_covariate}A), directly validating the targeted penalty design: excluding prior weights from $d^h_j$ ensures that prior biases do not suppress the detection of genuine context-specific perturbations. False-positive rates for null covariate effects remained consistently below 0.05 across all configurations
(Figure~\ref{fig:sim_covariate}B), with prior knowledge further reducing FPR variability across replicates, indicating more reliable model selection in real-world settings where reproducibility is essential.

Further results of the simulation study, including per-configuration numerical summaries across all 24 factorial configurations, are presented in Supplementary Section~S2.

\subsection{T2D protein interaction network analysis}
\label{subsec:t2d_results}

\begin{figure}[htbp]
\centering
\includegraphics[width=\columnwidth]{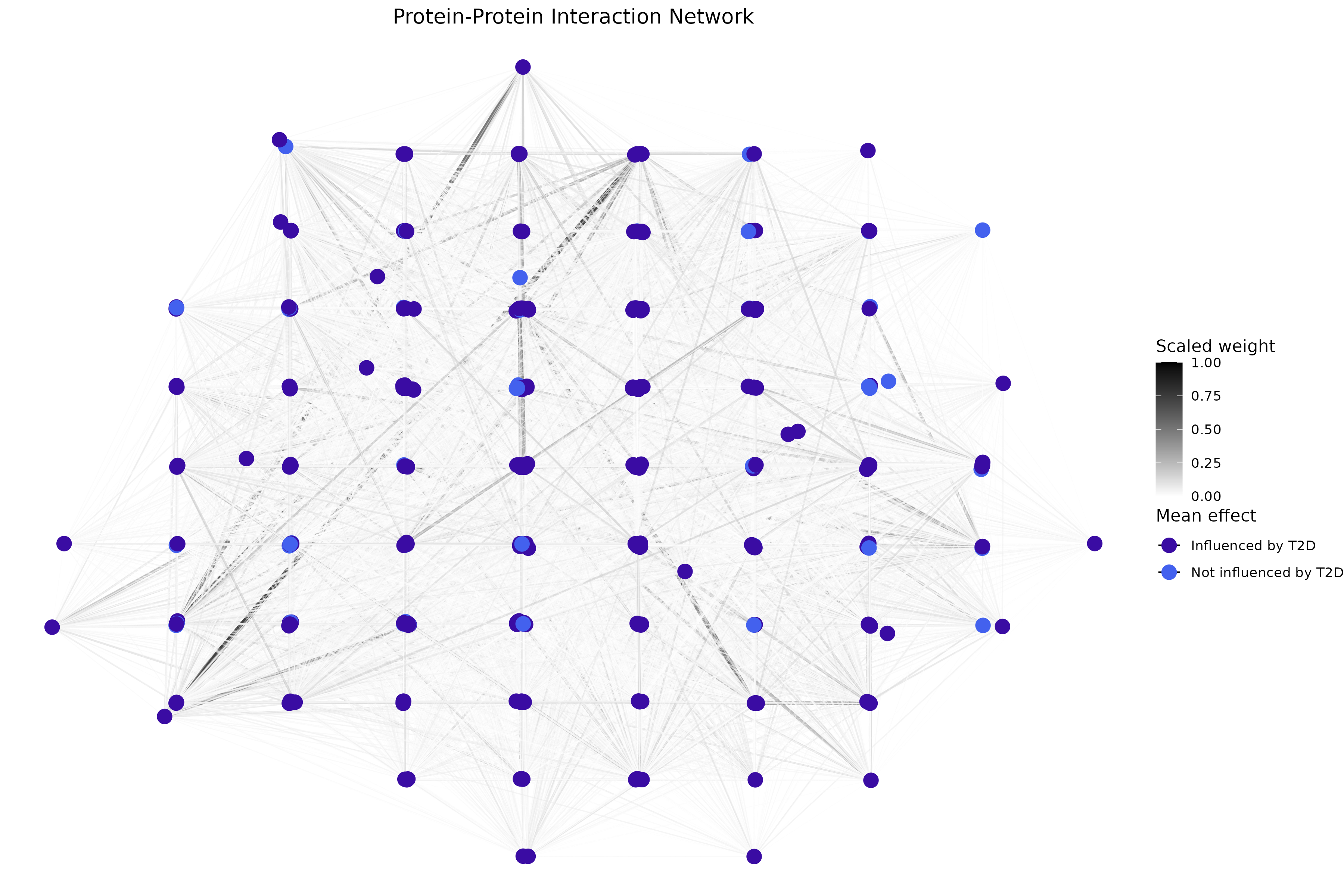}
\caption{T2D-associated protein interaction network. Full differential
network $\hat{\mathbf{A}}^{\mathrm{T2D}}$ estimated from the UK Biobank cardiometabolic proteomics data ($n=49{,}129$, $p=366$ proteins). Edge colour reflects the estimated partial correlation strength (darker: stronger associations). Node colour indicates T2D-dependent mean expression change.}
\label{fig:t2d_network}
\end{figure}

Results of the application to UK Biobank cardiometabolic proteomics data are shown in Figure~\ref{fig:t2d_network}. The reconstructed differential network $\hat{\mathbf{A}}^{\mathrm{T2D}}$ exhibited high density, with 98\% of protein pairs showing non-zero modulated associations, and the majority of proteins displayed T2D-dependent changes in mean expression. This pervasive signal is expected: T2D is a systemic metabolic disorder with well-established broad effects on cardiovascular and inflammatory protein networks~\citep{zheng2018global}, and the 366 profiled proteins were specifically selected for their cardiometabolic relevance, making widespread co-regulatory disruption under such a condition likely.

\paragraph{Network-central proteins and biomarker candidates.}
To identify proteins with biomarker potential, we computed five complementary centrality measures (degree, eigenvector, betweenness, PageRank, and hub score; formal definitions in Supplementary Section~S4) and retained proteins lying in the intersection of the top 10\% ranked by overall centrality and hub score, yielding 34 candidate biomarkers. Table~\ref{tab:top_proteins} reports a representative subset, while the complete list is provided in Table~S1 and Figure~\ref{fig:S1}. Most identified proteins (79\%) have previously documented associations with T2D in DisGeNET~\citep{pinero2016disgenet}, supporting the biological validity of the inferred network structure. Critically, several identified proteins, including CHEK2, CNDP1, and NTproBNP, lacked significant differential expression between non-diabetic and incident diabetic individuals, yet occupied central network positions (Table~\ref{tab:top_proteins}, rows 4-6). These proteins, invisible to conventional differential expression analysis, may represent regulatory nodes whose importance lies in their connectivity rather than in mean expression changes. Using the 34 central proteins, a T2D risk prediction model was trained on a balanced subset and tested on an independent cohort, achieving a test AUC of 63.2\% (95\% CI: 55.2--71.2\%) for predicting 5-year T2D incidence (Figure~\ref{fig:S3}). Further details about the T2D risk prediction model are available in Supplementary Section~S5.

\begin{table}[htbp]
\caption{Representative network-central proteins identified through
combined centrality analysis. \textbf{Bold}: proteins not differentially
expressed (novel network-based candidates). Columns indicate: protein gene name; known T2D association from DisGeNET database; DisGeNET gene-disease association score that takes into account the number of sources supporting the association and the reliability of each of them; differential expression status, and fold change values. Full table with all 34 proteins is provided in Table~S1.}
\label{tab:top_proteins}
\begin{tabular*}{\columnwidth}{@{\extracolsep\fill}lcccc@{\extracolsep\fill}}
\toprule
Gene & T2D in & DisGeNET & Diff.\ & Fold \\
     & DisGeNET & score    & expressed & change \\
\midrule
LEP       & Yes & 1.00 & Yes  & 1.15 \\
GH1       & Yes & 1.00 & Yes  & 0.86 \\
LPL       & Yes & 1.00 & Yes  & 0.94 \\
\textbf{CHEK2}     & Yes & 0.10 & \textbf{No}  & -- \\
\textbf{CNDP1}     & Yes & 0.80 & \textbf{No}  & -- \\
\textbf{NTproBNP}  & No  & --   & \textbf{No}  & -- \\
ACE2      & Yes & 0.90 & Yes  & 1.15 \\
FABP4     & Yes & 0.55 & Yes  & 1.14 \\
HMOX1     & Yes & 0.95 & Yes  & 1.00 \\
IGFBP1    & Yes & 0.55 & Yes  & 0.79 \\
\end{tabular*}
\end{table}

\begin{figure}[htbp]
\centering
\includegraphics[width=\columnwidth]{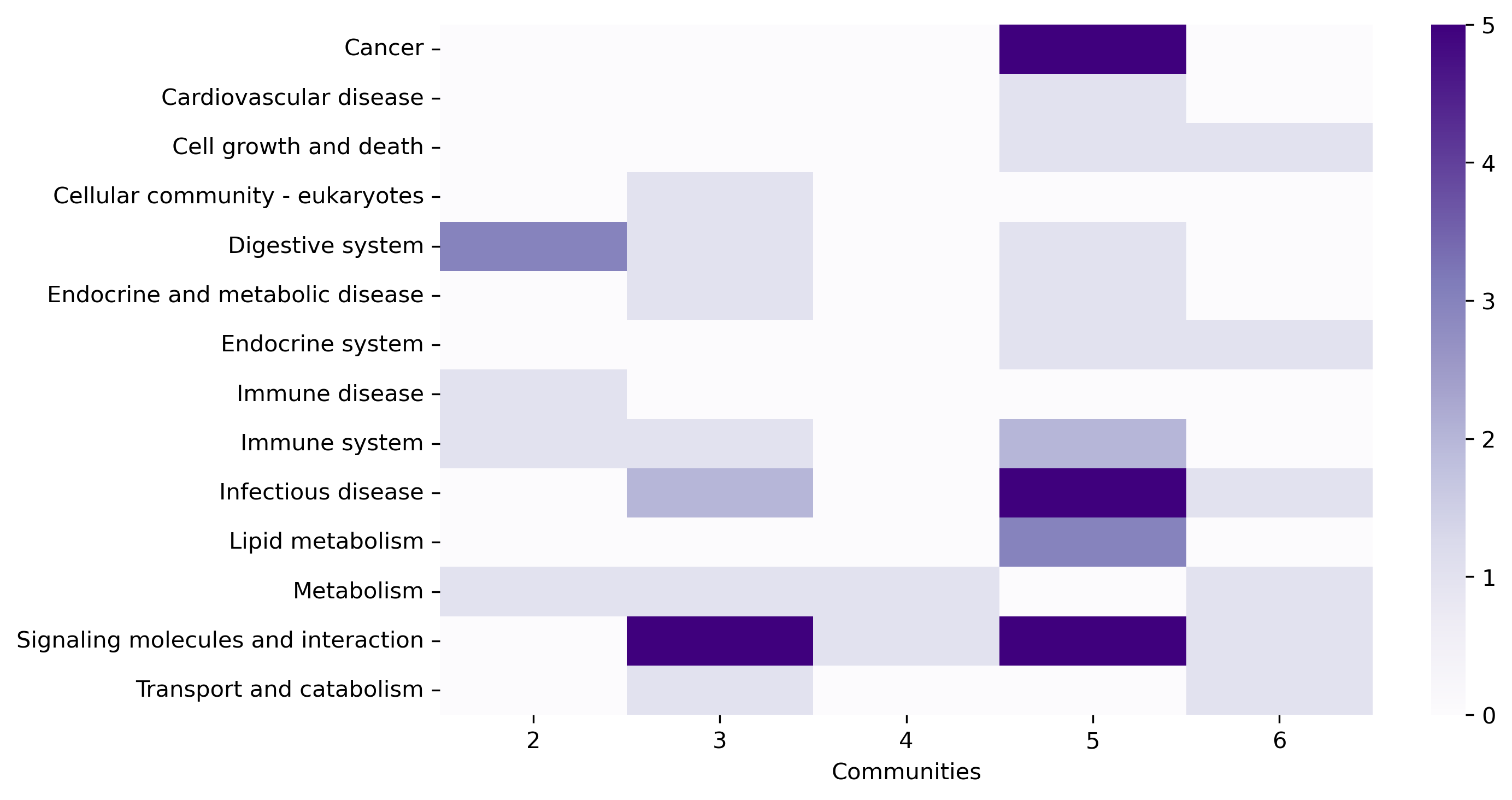}
\caption{KEGG pathway enrichment across T2D-associated protein
communities. Heatmap shows the number of significantly enriched pathways
(adjusted $p\leq 0.05$) per KEGG category for each of the six communities
identified by fast greedy modularity optimisation. Community~1 had no
significant enrichments. Colour intensity indicates pathway count
(scale 0--5+).}
\label{fig:enrichment}
\end{figure}

\paragraph{Community structure and functional organisation.}
Community detection via fast greedy modularity optimisation~\citep{clauset2004finding} identified six functionally distinct protein modules (Figure~\ref{fig:S2}; community membership in Table~S4). KEGG pathway enrichment~\citep{kanehisa2025kegg} revealed coherent biological specialisation across communities (Figure~\ref{fig:enrichment}). Community~5 showed the strongest enrichment profile, with maximal representation of cancer pathways and significant cardiovascular disease associations, reflecting the well-established metabolic-oncogenic-cardiovascular co-morbidity axis in T2D. Community~3 emerged as a broad metabolic hub spanning endocrine, immune, and cellular transport pathways, suggesting a core regulatory module integrating multiple physiological systems affected by diabetes. Community~2 was specifically enriched for digestive system pathways, consistent with gastrointestinal manifestations of T2D. Most communities showed significant enrichment for signalling molecules and interaction pathways, indicating widespread disruption of intercellular communication networks. The clear functional segregation across modules confirms that the T2D-associated network perturbations occur within coherent biological programmes rather than randomly across the proteome.

%% file: Discussion.tex
Biological networks constructed from omics data offer a systems-level window into disease mechanisms that single-molecule analyses cannot provide. Among molecular layers, protein-protein interaction networks are particularly valuable: proteins constitute the dominant drug-target class in precision medicine, and the availability of large-scale curated databases makes them ideally suited to the prior-informed framework introduced here. In order to construct biological networks more reliable for downstream analysis, we have introduced a prior-informed conditional graphical model that jointly addresses two limitations in the field of GGM-based biological network estimation: the omission of curated prior knowledge and the inability to model individual-level network heterogeneity. The core contribution, a structured weighted penalty applied selectively to the population-level estimation while leaving covariate-specific effects entirely data-driven, yields consistent improvements across simulated settings. A practically important finding is the method's robustness to prior imperfection: STRING scores and similar databases are inevitably incomplete and biased towards well-characterised proteins, yet incorporating even noisy prior knowledge substantially improves network reconstruction without suppressing novel discoveries, broadening applicability to contexts where comprehensive databases are sparse or unavailable. The T2D case study exemplifies the translational potential of this approach, successfully identifying both established diabetes-associated proteins and novel network-central biomarkers that would be missed by conventional differential expression analysis (CHEK2, CNDP1, NTproBNP). The discovery of six distinct protein communities with plausible functional specialisation, ranging from cancer and signalling pathways to immune and metabolic modules, provides systems-level insights into diabetes pathogenesis while demonstrating the biological interpretability of network-derived results. The methodology enables identification of both individual biomarkers, identifying key regulatory nodes, and functional modules that are perturbed in disease contexts~\citep{ma2019differential,mousavian2019differential}, providing complementary perspectives on disease mechanisms.
These results demonstrate the model's ability to capture meaningful, covariate-modulated network structures, providing a principled basis for patient-specific interaction structure estimation that can inform personalised risk stratification and precision medicine strategies based, for example, on patient-specific network vulnerabilities~\citep{zhang2025resilience}. The R implementation, accessible via a single function with no mandatory inputs beyond expression data, transforms complex statistical methodology into accessible tools for the broader biomedical community.

The framework carries limitations that should inform its application. The Gaussian assumption and linear parameterisation of covariate effects are reasonable for many omics applications but may be challenged by heavy-tailed distributions, multimodal expression patterns, or nonlinear covariate relationships; preprocessing approaches such as nonparanormal transformations~\citep{liu2009nonparanormal} can mitigate distributional violations. Sample size requirements scale with model complexity: while the simulations confirmed reliable performance at $n=500$, applications with many covariates or high-dimensional networks may require larger cohorts, potentially limiting applicability in smaller specialised studies. Finally, as with all neighbourhood selection approaches, the node-wise regression framework may be less optimal for highly dense or globally structured networks, where joint estimation methods may better capture global dependencies.

These constraints point to natural extensions: nonlinear covariate modelling to capture threshold effects and complex regulatory relationships, multi-omics generalisation to extend the model from protein to gene and metabolite networks, and network-aware risk scoring that weights biomarker contributions by their position within the personalised interaction structure. Prior-informed conditional GGMs offer a principled starting point for the next generation of personalised network analyses in precision medicine.

%% file: Supplementary.tex

\clearpage

\setcounter{figure}{0}
\setcounter{table}{0}
\renewcommand{\thefigure}{S\arabic{figure}}
\renewcommand{\thetable}{S\arabic{table}}


\begin{center}
{\large\textbf{Supplementary Materials}}\\[0.5em]
{\normalsize Prior-informed conditional Gaussian graphical models: an application to protein interaction network reconstruction.}\\[0.3em]
{\small Alessia Mapelli, Michela Carlotta Massi, Gianmauro Cuccuru,
Emanuele Di Angelantonio, Francesca Ieva}
\end{center}

\bigskip


\subsection*{Supplementary section S1: methodological and implementation details}
\label{supp:met_details}

This section provides technical details complementing the Methods section
of the main text. We use $j \in \{1,\ldots,p\}$ as the generic node index
throughout; the main text sets $j=p$ without loss of generality for
notational convenience.

\subsubsection*{S1.1\quad Covariate preprocessing}

Proper covariate preprocessing is essential for both numerical stability
and interpretability of the estimated network effects. Two encoding
strategies are applied depending on the covariate type.

\textit{Continuous covariates} are standardised to zero mean and unit
variance before estimation. Under this encoding, the baseline precision
matrix $\boldsymbol{\Delta}^0$ reflects population network structure at the
covariate mean. Each $\boldsymbol{\Delta}^h$ captures the change in
network structure associated with a one-standard-deviation increase in
covariate $h$, with the scale of $\hat{\beta}_j$ and
$\hat{\delta}^h_{jk}$ directly comparable across covariates.

\textit{Categorical covariates} with $C$ levels are encoded via $C-1$
binary dummy variables, with the most frequent category as the reference.
Under this encoding, $\boldsymbol{\Delta}^0$ reflects the network structure
of the reference group, and each $\boldsymbol{\Delta}^h$ captures the
deviation in network structure for category $h$ relative to that reference.

\subsubsection*{S1.2\quad Step 1: mean parameter estimation}

For each node $j$, the covariate effect on the protein mean is estimated
by solving the $\ell_1$-penalised regression:
\begin{equation}
\hat{\beta}_j = \arg\min_{\beta \in \mathbb{R}^q}\;
  \frac{1}{2n}\bigl\|Y_j - X\beta\bigr\|_2^2 + \lambda_j^B\|\beta\|_1,
\label{eq:S_step1}
\end{equation}
where $X \in \mathbb{R}^{n \times q}$ is the (preprocessed) design matrix.
The regularisation parameter $\lambda_j^B$ is selected by $K$-fold
cross-validation ($K=5$ in all experiments) using the \texttt{glmnet}
package~\citep{friedman2010regularization}. The mean effect matrix is constructed as $\mathbf{B}_{j\cdot} = \hat{\beta}_j^T$.

The mean-regression residuals $Z_j = Y_j - X\hat{\beta}_j$ are then
passed to Step~2.

\subsubsection*{S1.3\quad Step 2: prior-informed neighborhood estimation}

\paragraph{Residual regression formulation.}
After mean removal, the conditional model in Eq.~\ref{eq:neigh_cond_eq}
of the main text reduces, for node $j$, to:
\begin{equation}
Z_j = \sum_{k \neq j} \delta^0_{jk} Z_k
      + \sum_{k \neq j}\sum_{h=1}^{q} \delta^h_{jk}(X_h Z_k) + \varepsilon_j,
\quad \varepsilon_j \sim \mathcal{N}(0, \sigma_{jj}^{-1}).
\label{eq:S_resid}
\end{equation}
The full parameter vector for node $j$ is
$d_j = [(d^0_j)^T, (d^1_j)^T, \ldots, (d^q_j)^T]^T \in \mathbb{R}^{(q+1)(p-1)}$,
where $d^0_j = (\delta^0_{j,k})_{k \neq j}$ collects the $(p-1)$
population-level coefficients and $d^h_j = (\delta^h_{j,k})_{k \neq j}$
collects the $(p-1)$ coefficients for covariate $h$.

\paragraph{Design matrix structure.}
The predictor matrix $\mathbf{U}_{-j} \in \mathbb{R}^{n \times (q+1)(p-1)}$
is constructed by column-binding the residuals of the other nodes and their
covariate interactions:
\begin{equation}
\mathbf{U}_{-j} =
  \bigl[\,Z_{-j},\;
        X_1 \odot Z_{-j},\;
        X_2 \odot Z_{-j},\;
        \ldots,\;
        X_q \odot Z_{-j}\,\bigr],
\label{eq:S_design}
\end{equation}
where $X_h \odot Z_{-j}$ denotes row-wise scaling of the residual matrix
$Z_{-j} \in \mathbb{R}^{n \times (p-1)}$ by the $n$-dimensional covariate
vector $X_h$. The first block contains the mean-centred residuals of all
nodes except $j$, encoding population-level associations; each subsequent
block captures how covariate $h$ modulates those associations through the
interaction terms $X_h Z_k$.

\paragraph{Estimation problem.}
Given the residuals $Z$ and the design matrix $\mathbf{U}_{-j}$, the
baseline and covariate-specific neighbourhood coefficients for node $j$
are jointly estimated by solving the prior-informed sparse-group Lasso:
\begin{equation}
\hat{d}_j = \arg\min_{d_j}\;
  \frac{1}{2n}\bigl\|Z_j - \mathbf{U}_{-j}d_j\bigr\|_2^2 \notag+ \alpha\lambda_j^\Delta
     \bigl\|(1-\mathbf{W}_{j,-j})\odot d^0_j\bigr\|_1 \label{eq:S_main}
+\vartheta\alpha\lambda_j^\Delta \bigl\|d^{-0}_j\bigr\|_1 + (1-\alpha)\lambda_j^\Delta \sum_{h=1}^{q}\bigl\|d^h_j\bigr\|_2, \notag
\end{equation}
where $d^{-0}_j = [(d^1_j)^T, \ldots, (d^q_j)^T]^T$ collects all
covariate-specific coefficients. The three penalty terms encode distinct
structural assumptions: (i) a prior-weighted $\ell_1$ penalty on the
baseline neighbourhood $d^0_j$, reducing shrinkage on database-supported
edges; (ii) a uniform $\ell_1$ penalty on all covariate-specific
coefficients $d^{-0}_j$, promoting element-wise sparsity; and (iii) a
group $\ell_2$ penalty on each covariate block $d^h_j$, encouraging all
edges modulated by a given covariate to be jointly absent or present.

\paragraph{Bayesian interpretation of the weighted penalty.}
The $\ell_1$ penalty in standard Lasso corresponds, under a Bayesian
interpretation, to independent Laplacian priors on regression coefficients:
$p(\delta_{jk}) \propto \exp(-\lambda|\delta_{jk}|)$~\citep{wang2013incorporating}. The weighted
neighbourhood selection extends this to a knowledge-informed edge-specific prior.
Specifically, the weighted $\ell_1$ penalty on $d^0_j$,
\begin{equation}
\alpha\lambda_j^\Delta \bigl\|(1-\mathbf{W}_{j,-j})\odot d^0_j\bigr\|_1
= \alpha\lambda_j^\Delta \sum_{k \neq j}(1-W_{jk})|\delta^0_{jk}|,
\end{equation}
corresponds to independent Laplacian priors with edge-specific scale:
$p(\delta^0_{jk}) \propto \exp\!\bigl(-\alpha\lambda_j^\Delta(1-W_{jk})
|\delta^0_{jk}|\bigr)$. When $W_{jk} \approx 1$ (strong prior support), the prior scale diverges and the coefficient faces almost no shrinkage; when $W_{jk} = 0$ (no prior support), the prior reduces to a standard
Laplacian with scale $\alpha\lambda_j^\Delta$.

For the covariate-specific effects, the element-wise penalty
$\vartheta\alpha\lambda_j^\Delta\|d^{-0}_j\|_1$ corresponds to
independent Laplacian priors with uniform scale $\vartheta\alpha\lambda_j^\Delta$,
reflecting the absence of prior knowledge about which edges are modulated
by context-specific factors. The group penalty
$(1-\alpha)\lambda_j^\Delta\sum_h\|d^h_j\|_2$ places a spherical
Gaussian-type prior on the vector of effects of each covariate
$h$, encouraging all edges modulated by a given
covariate to be jointly zero or non-zero,  a natural sparsity structure
when selecting influential covariates.

\paragraph{Optimisation and hyperparameter selection.}
The optimisation problem (Eq.~\ref{eq:main})  is solved using the
\texttt{sparsegl} package~\citep{liang2024sparsegl}, which implements an efficient majorized block-wise coordinate descent algorithm for sparse-group Lasso problems with
group-specific penalty weights. The per-node regularisation parameter $\lambda_j^\Delta$ is selected by $K$-fold cross-validation.

The structural parameters $\alpha \in [0,1]$ and $\vartheta > 0$ are
treated as tuning parameters and selected by minimising the Bayesian
Information Criterion summed over all nodes:
\begin{equation}
\text{BIC}(\alpha, \vartheta) =
  \sum_{j=1}^{p}
  \left[n \log\!\left(\frac{\text{RSS}_j}{n}\right)
        + \log(n)\cdot\mathrm{df}_j\right],
\label{eq:S_bic}
\end{equation}
where $\text{RSS}_j = \|Z_j - \mathbf{U}_{-j}\hat{d}_j\|_2^2$ and
$\mathrm{df}_j = \|\hat{d}_j\|_0$ is the number of non-zero estimated
coefficients for node $j$. Candidate values are
$\alpha \in \{0.50, 0.75, 0.90, 0.99\}$, spanning from predominantly
element-wise penalties ($\alpha \to 1$) to predominantly group penalties
($\alpha \to 0$), and $\vartheta \in \{0.80, 1.00, 1.10, 1.30\}$,
controlling the relative stringency of the covariate-effect penalty
relative to the prior-informed baseline penalty. For larger parameter spaces, a grid-based random search is supported as an alternative to exhaustive enumeration.

\paragraph{Pre-screening.}
When $p \times q \gg n$, the $(q+1)(p-1)$-dimensional optimisation in Eq.\ref{eq:main} can be computationally demanding. To reduce problem dimensionality, a pre-screening step can be applied for each node $j$, excluding predictor $k$ from node $j$'s regression based on one of two
strategies. In the \emph{correlation-based} strategy, predictor $k$ is excluded if $W_{jk} = 0$ (no prior support) and
$|\mathrm{cor}(Z_j, Z_k)| < \gamma$, where $\gamma$ is a data-driven threshold set to the $r$-th percentile ($r = 20$ by default) of all $\binom{p}{2}$ absolute pairwise correlations. Alternatively, the user may supply a \emph{custom screening matrix} $S \in \{0,1\}^{p \times p}$, where
$S_{jk} = 0$ indicates that predictor $k$ should be excluded from node $j$'s regression; this allows domain knowledge to directly drive feature pre-selection. In both strategies, pre-screening is never applied when $W_{jk} > 0$, ensuring
that prior-supported interactions are always evaluated regardless of empirical correlation or the entries of $S$. This procedure is analogous to the sure independence screening principle\citep{fan2008sure} and is particularly effective in biological applications where most node pairs have no direct interaction. Pre-screening is available as an option in the
implemented software but was not applied in any of the analyses reported in this paper.

\subsubsection*{S1.4\quad Step 3: symmetrisation and network reconstruction}
For any $h=0,\dots,q$, node-wise regressions yield asymmetric estimates: $\hat{\delta}^h_{jk}$
(obtained from node $j$'s regression) and $\hat{\delta}^h_{kj}$ (from
node $k$'s regression) are usually not be equal. Two symmetrisation strategies
are available, trading off network density against specificity.

The \textbf{OR rule} (used throughout the main analyses) retains edge
$(j,k)$ in $E^h$ if either directional estimate is
non-zero:
\begin{equation}
E^h = \{(p,k): \delta^h_{p,k} \neq 0\ | \delta^h_{k,p} \neq 0\}
\end{equation}
The edge weights are then set to the coefficient with larger absolute value, representing an estimate of the partial correlation between $j$ and $k$ attributable to context $h$:
\begin{equation}
  \hat{\rho}^h_{pk|-\{p,k\}} =  \hat{\delta}^h_{p,k}\mathds{1}_{|\hat{\delta}^h_{p,k}| \geq |\hat{\delta}^h_{k,p}|} + \hat{\delta}^h_{k,p}\mathds{1}_{|\hat{\delta}^h_{p,k}| < |\hat{\delta}^h_{k,p}|}
\end{equation}
The symmetrised precision matrix components $\hat{\boldsymbol{\Delta}}^h$ are then constructed as:
\begin{align}
[\hat{\boldsymbol{\Delta}}^h]_{jk} &= [\hat{\boldsymbol{\Delta}}^h]_{kj}
\nonumber \\
&= -\!\left(
  \hat{\sigma}_{jj}\hat{\delta}^h_{jk}
    \,\mathds{1}_{\{|\hat{\delta}^h_{jk}|\,\geq\,|\hat{\delta}^h_{kj}|\}}
  + \hat{\sigma}_{kk}\hat{\delta}^h_{kj}
    \,\mathds{1}_{\{|\hat{\delta}^h_{jk}|\,<\,|\hat{\delta}^h_{kj}|\}}
\right),
\label{eq:S_or}
\end{align}
where $\hat{\sigma}_{jj} = 1/\widehat{\mathrm{Var}}(\varepsilon_j)$ is
estimated from the node-$j$ regression residuals.

The \textbf{AND rule} is more conservative, retaining edge $(j,k)$ only
when both directional estimates are non-zero:
\begin{equation}
E^h = \{(p,k): \delta^h_{p,k} \neq 0 \And \delta^h_{k,p} \neq 0\}
\end{equation}
and selecting the coefficient with smaller absolute value as weight:
\begin{equation}
  \hat{\rho}^h_{pk|-\{p,k\}} =  \hat{\delta}^h_{p,k}\mathds{1}_{|\hat{\delta}^h_{p,k}| < |\hat{\delta}^h_{k,p}|} + \hat{\delta}^h_{k,p}\mathds{1}_{|\hat{\delta}^h_{p,k}| \geq |\hat{\delta}^h_{k,p}|}
\end{equation}
The symmetrised precision matrix components are then constructed as:
\begin{align}
[\hat{\boldsymbol{\Delta}}^h]_{jk} &= [\hat{\boldsymbol{\Delta}}^h]_{kj}
\nonumber \\
&= -\!\left(
  \hat{\sigma}_{jj}\hat{\delta}^h_{jk}
    \,\mathds{1}_{\{|\hat{\delta}^h_{jk}|\,<\,|\hat{\delta}^h_{kj}|\}}
  + \hat{\sigma}_{kk}\hat{\delta}^h_{kj}
    \,\mathds{1}_{\{|\hat{\delta}^h_{jk}|\,\geq\,|\hat{\delta}^h_{kj}|\}}
\right).
\label{eq:S_and}
\end{align}
The OR rule produces denser networks and is preferable when the goal is
broad discovery (e.g.\ biomarker screening); the AND rule imposes higher
confidence thresholds and is preferable when controlling false positives
is of prime importance.

\paragraph{Personalised network prediction.}
Thanks to the linear parameterisation in Eq.~\ref{eq:conditional}, the personalised adjacency matrix for a new individual with
covariate vector $x$ is obtained by direct linear
combination of the estimated components:
\begin{equation}
\hat{\mathbf{A}}(x)_{jk}
  = \hat{\rho}^0_{jk|\text{-}\{j,k\}}
  + \sum_{h=1}^{q} x_h\,\hat{\rho}^h_{jk|\text{-}\{j,k\}}.
\label{eq:S_personalized}
\end{equation}
Covariates in $x$ must be preprocessed using the same standardisation
parameters estimated from the training set, and categorical variables
must be encoded consistently with the training encoding. The
\texttt{predict\_personalized\_network} function in the software package
handles this preprocessing automatically.


\subsection*{Supplementary section S2: additional simulation results}
\label{supp:add_sim_results}

\paragraph{Simulation configurations.}
Table~\ref{tab:sim_configs} summarises the full factorial simulation design
(24 unique configurations, each replicated 10 times).
Performance metrics are averaged over all replicates within each configuration.

\begin{table*}[htbp]
\small
\caption{Simulation study: per-configuration performance summary across
all 24 factorial configurations (10 replicates each, 240 runs total).
Acc: accuracy; F1: F1-score; MagCor: Spearman correlation between
estimated and true precision matrix entries,
all computed on the baseline component $\boldsymbol{\Delta}^0$.
Cov.\ FPR: false positive rate for covariate-specific effects; for
$q=1$ this is the FPR of the single covariate; for $q=3$ this is the
maximum FPR across the three covariates (conservative bound).
Time: mean computational time per node (minutes), averaged over
replicates. All metrics are means over 10 replicates.}
\label{tab:sim_configs}
\begin{tabular*}{\textwidth}{@{\extracolsep\fill}ccccccc@{\extracolsep\fill}}
\toprule
$n$ & $p$ & $q$ & Prior & Baseline (Acc / F1 / MagCor) & Cov.\ FPR & Time (min) \\
\midrule
500  & 50  & 1 & Perfect & 0.978 / 0.777 / 0.799 & 0.021 & 3.3 \\
500  & 50  & 1 & Noisy   & 0.945 / 0.597 / 0.671 & 0.028 & 4.2 \\
500  & 50  & 1 & None    & 0.904 / 0.456 / 0.569 & 0.048 & 1.6 \\
500  & 50  & 3 & Perfect & 0.982 / 0.812 / 0.829 & 0.012 & 9.7 \\
500  & 50  & 3 & Noisy   & 0.956 / 0.648 / 0.707 & 0.014 & 12.0 \\
500  & 50  & 3 & None    & 0.910 / 0.469 / 0.575 & 0.033 & 4.0 \\
500  & 100 & 1 & Perfect & 0.986 / 0.729 / 0.751 & 0.011 & 14.7 \\
500  & 100 & 1 & Noisy   & 0.970 / 0.575 / 0.643 & 0.011 & 17.9 \\
500  & 100 & 1 & None    & 0.943 / 0.407 / 0.514 & 0.019 & 7.0 \\
500  & 100 & 3 & Perfect & 0.993 / 0.840 / 0.848 & 0.006 & 15.6 \\
500  & 100 & 3 & Noisy   & 0.973 / 0.598 / 0.661 & 0.005 & 11.6 \\
500  & 100 & 3 & None    & 0.946 / 0.424 / 0.528 & 0.014 & 33.7 \\
1000 & 50  & 1 & Perfect & 0.966 / 0.706 / 0.748 & 0.031 & 3.8 \\
1000 & 50  & 1 & Noisy   & 0.938 / 0.566 / 0.649 & 0.038 & 4.5 \\
1000 & 50  & 1 & None    & 0.876 / 0.396 / 0.529 & 0.068 & 1.9 \\
1000 & 50  & 3 & Perfect & 0.977 / 0.774 / 0.801 & 0.022 & 9.1 \\
1000 & 50  & 3 & Noisy   & 0.943 / 0.593 / 0.668 & 0.019 & 10.2 \\
1000 & 50  & 3 & None    & 0.892 / 0.426 / 0.550 & 0.040 & 3.9 \\
1000 & 100 & 1 & Perfect & 0.988 / 0.777 / 0.799 & 0.011 & 14.6 \\
1000 & 100 & 1 & Noisy   & 0.966 / 0.546 / 0.621 & 0.014 & 17.9 \\
1000 & 100 & 1 & None    & 0.932 / 0.372 / 0.493 & 0.029 & 6.9 \\
1000 & 100 & 3 & Perfect & 0.993 / 0.842 / 0.855 & 0.008 & 21.8 \\
1000 & 100 & 3 & Noisy   & 0.974 / 0.609 / 0.670 & 0.006 & 21.5 \\
1000 & 100 & 3 & None    & 0.942 / 0.407 / 0.518 & 0.017 & 17.6 \\
\end{tabular*}
\end{table*}

\paragraph{Supplementary simulation observations.}
Across the full factorial design, the perfect prior uniformly
outperformed the noisy prior, which, in turn, uniformly outperformed the no prior across all metrics and configuration combinations. The most pronounced improvements appear in F1-score and magnitude correlation (MagCor), which are more informative than accuracy in sparse networks where the majority of edges are
absent by construction. As illustrative extremes, at $n=1{,}000$, $p=100$,
$q=3$, the perfect prior raises F1 from 0.407 (no prior) to 0.842 and MagCor from 0.518 to 0.855; even the noisy prior, which has 20\% false negatives and 10\% false positives, achieves F1$\,=\,$0.609 and MagCor$\,=\,$0.670, already a
substantial improvement over no-prior. The prior benefit is maintained or slightly
amplified at $p=100$ relative to $p=50$, consistent with greater regularisation being more consequential in higher-dimensional settings.

Covariate-specific estimation is unaffected by prior knowledge across all
configurations, as intended: the false positive rate for covariate effects remains
below 0.07 in all cases and below 0.04 whenever any prior information is
available, confirming that the weighted penalty does not bleed into the covariate-specific components. In the $q=3$ scenario, the maximum FPR across
covariates stays below 0.04 for all prior conditions, demonstrating reliable
covariate selection even under the added complexity of multiple simultaneous
network modulations.


\subsection*{Supplementary section S3: UK Biobank data and preprocessing details}
\label{supp:UKB_preprocessing}

\paragraph{Study population and cohort construction.}
We applied the framework to data from the UK Biobank Plasma Proteome Project (UKB-PPP)~\citep{sun2023plasma}, which provides large-scale proteomic profiling alongside detailed phenotypic information for a population-based prospective cohort. The UK Biobank recruited approximately 500,000 participants aged 40–69 years, with long-term follow-up through national health registries. Within this cohort, proteomics
measurements were obtained using the Olink Explore 3072 platform on plasma samples from 54{,}219 randomly selected participants, capturing 2{,}923
unique proteins across eight panels. For this T2D-focused analysis, we
concentrated on proteins from the Olink Cardiometabolic I and II panels,
applying quality control procedures following established
protocols~\citep{sun2023plasma}.

Quality control was performed in the following order: (1) proteins with $>10\%$ missing values were excluded; (2) samples with $>50\%$ missing protein measurements were removed; (3) mean-based imputation was applied to remaining missing values. After QC, 366 proteins were retained across
50{,}531 samples.

\paragraph{Analytical cohort definition and T2D outcome ascertainment.}
A healthy baseline population was established using the following exclusion criteria applied at the proteomics assessment date:
\begin{itemize}
\item Prior type 1 diabetes (T1D) diagnosis (ICD-10: E10);
\item Prior type 2 diabetes (T2D) diagnosis (ICD-10: E11--E14);
\item Antidiabetic medication use at baseline;
\item HbA1c measurements $>48\,\mathrm{mmol/mol}$ at baseline.
\end{itemize}
This yielded an analytical cohort of 49{,}129 eligible individuals.
Incident T2D within 5 years of proteomics assessment was determined through any of the following: ICD-10 codes E10-E14 from death registries, primary care records, or hospital admissions; self-reported
diabetes at follow-up; HbA1c $>48\,\mathrm{mmol/mol}$ at follow-up; or antidiabetic drug prescriptions.
\paragraph{Cohort partitioning.}
The 49{,}129-individual cohort was split into three independent subsets
(Table~\ref{tab:cohort}):
\begin{enumerate}
\item \textbf{Estimation cohort} (80\%, $n=39{,}304$): used for network
reconstruction, preserving the observed T2D incidence and sex distribution
through stratified sampling;
\item \textbf{Biomarker training set} ($n=270$): a more balanced subset with one incident T2D case and four controls, matched by sex and age, used for logistic regression biomarker model development;
\item \textbf{Test cohort} ($n=9{,}555$): the remaining individuals used for
independent evaluation of the T2D risk prediction model.
\end{enumerate}

\paragraph{Covariate selection and preprocessing.}
Covariate selection was guided by the validated QDiabetes-2018 clinical risk framework~\citep{hippisley2017development}. We assessed the covariate
associations with protein expression through principal component analysis
of the proteomic data, identifying sex as the dominant determinant of
proteomic variation ($R^2 \approx 50\%$ for PC6). Based on this analysis
and the clinical focus of the study, two covariates were included:
\begin{itemize}
\item \textbf{Sex}: binary indicator (0: female, 1: male), treated as a confounding covariate modelling systematic sex-differences in PPI network
structure;
\item \textbf{T2D incidence}: binary indicator (0: non-diabetic, 1:
incident T2D), the primary covariate of interest for identifying
T2D-specific network perturbations.
\end{itemize}
Continuous protein expression values were standardised (zero mean, unit
variance) prior to network estimation.

\begin{table}[htbp]
\caption{UK Biobank analytical cohort partitioning.
T2D\,Inc.: 5-year incident T2D; F: female; M: male.}
\label{tab:cohort}
\begin{tabular*}{\columnwidth}{@{\extracolsep\fill}lcccc@{\extracolsep\fill}}
\toprule
Subset & $n$ & T2D\,Inc.\ (\%) & Age (mean\,$\pm$\,sd) & Sex (F/M) \\
\midrule
Estimation    & 39,304 & 1.1 & 56.59 $\pm$ 8.23 & 21622/17682 \\
Training (balanced) & 270 & 20.0 & 60.39 $\pm$ 6.85 & 120/150 \\
Test          & 9,555  & 0.5 & 56.54 $\pm$ 8.25 & 5305/4250 \\
\midrule
\textbf{Total} & \textbf{49,129} & \textbf{1.1} & \textbf{56.60 $\pm$ 8.23} & \textbf{27047/22082} \\
\end{tabular*}
\end{table}

\paragraph{Prior knowledge extraction and processing.}
The prior weight matrix $\mathbf{W}$ was constructed from STRING database (v12.0) combined scores $s_{jk} \in [0, 1000]$ via $W_{jk} = s_{jk}/1000$, mapping
scores to $[0,1]$. Pairs absent from the database are
assigned $W_{jk} = 0$, receiving the strongest penalty and treated as
having no prior support. Pairs with $W_{jk} = 1$ receive no 
penalty, indicating that the prior fully trusts the interaction.

\paragraph{Network estimation hyperparameters.}
The network was estimated using the proposed prior-informed conditional GGM framework with the following protocol:
\begin{itemize}
\item Regularisation parameters $\lambda_j^B$ and $\lambda_j^\Delta$:
  5-fold cross-validation per node;
\item Structural parameters: random search over
  $\alpha \in \{0.50, 0.75, 0.90, 0.95\}$ and
  $\vartheta \in \{0.80, 1.00, 1.10, 1.30, 1.50\}$
  with BIC model selection;
\item Symmetrisation: OR rule (more inclusive, appropriate for dense
  biological networks);
\item Computational deployment: SLURM array jobs (one job per node,
  366 independent jobs on HPC infrastructure).
\end{itemize}


\subsection*{Supplementary section S4: network centrality measures}
\label{supp:centrality}

The following centrality measures were computed on the T2D-specific differential network $\hat{\mathbf{A}}^{\mathrm{T2D}}$, where edge weights correspond to estimated differential partial correlations. Let $p$ denote the number of nodes and $\hat{A}^{\mathrm{T2D}}_{jk}$ the edge weight between nodes $j$ and $k$.

\textbf{Degree centrality.} The weighted degree of node $j$ is the sum of absolute edge weights:
\begin{equation}
C_D(j) = \sum_{k=1}^p \bigl|\hat{A}^{\mathrm{T2D}}_{jk}\bigr|.
\end{equation}

\textbf{Eigenvector centrality.} The eigenvector centrality $C_E(j)$ satisfies the self-consistency equation
\begin{equation}
C_E(j) = \frac{1}{\lambda_1} \sum_{k=1}^p \hat{A}^{\mathrm{T2D}}_{jk}\,C_E(k),
\end{equation}
where $\lambda_1$ is the largest eigenvalue of $\hat{\mathbf{A}}^{\mathrm{T2D}}$~\citep{bonacich1987power}. Nodes with high eigenvector centrality are connected to other highly central nodes.

\textbf{Betweenness centrality.} The betweenness of node $k$ counts the fraction of shortest paths between all node pairs $(s,t)$ that pass through $k$:
\begin{equation}
C_B(k) = \sum_{s \neq k \neq t} \frac{\sigma_{st}(k)}{\sigma_{st}},
\end{equation}
where $\sigma_{st}$ is the total number of shortest paths from $s$ to $t$ and $\sigma_{st}(k)$ is the number of those paths passing through $k$.

\textbf{PageRank centrality.} PageRank is defined recursively as
\begin{equation}
C_{PR}(j) = \frac{1-d}{p} + d \sum_{k:\,\hat{A}^{\mathrm{T2D}}_{kj} \neq 0} \frac{C_{PR}(k)}{C_D(k)},
\end{equation}
with damping parameter $d = 0.85$~\citep{brin1998anatomy}. It captures influence through random-walk dynamics, upweighting nodes that receive links from other influential nodes.

\textbf{Hub score.} Hub scores are derived from the HITS (Hyperlink-Induced Topic Search) algorithm~\citep{kleinberg1999authoritative}, which simultaneously computes hub and authority scores via the dominant singular vectors of the adjacency matrix. Nodes with high hub scores connect to many high-authority nodes in the network.

Candidate biomarker proteins were defined as the intersection of the top-10\% nodes ranked by the mean of $C_D$, $C_E$, $C_B$, and $C_{PR}$, and the top-10\% nodes ranked by hub score, yielding 34 proteins (Table~S3).


\subsection*{Supplementary section S5: T2D risk prediction model details}
\label{supp:T2D_pred_model}

\begin{figure}[htbp]
\centering
\includegraphics[width=\columnwidth]{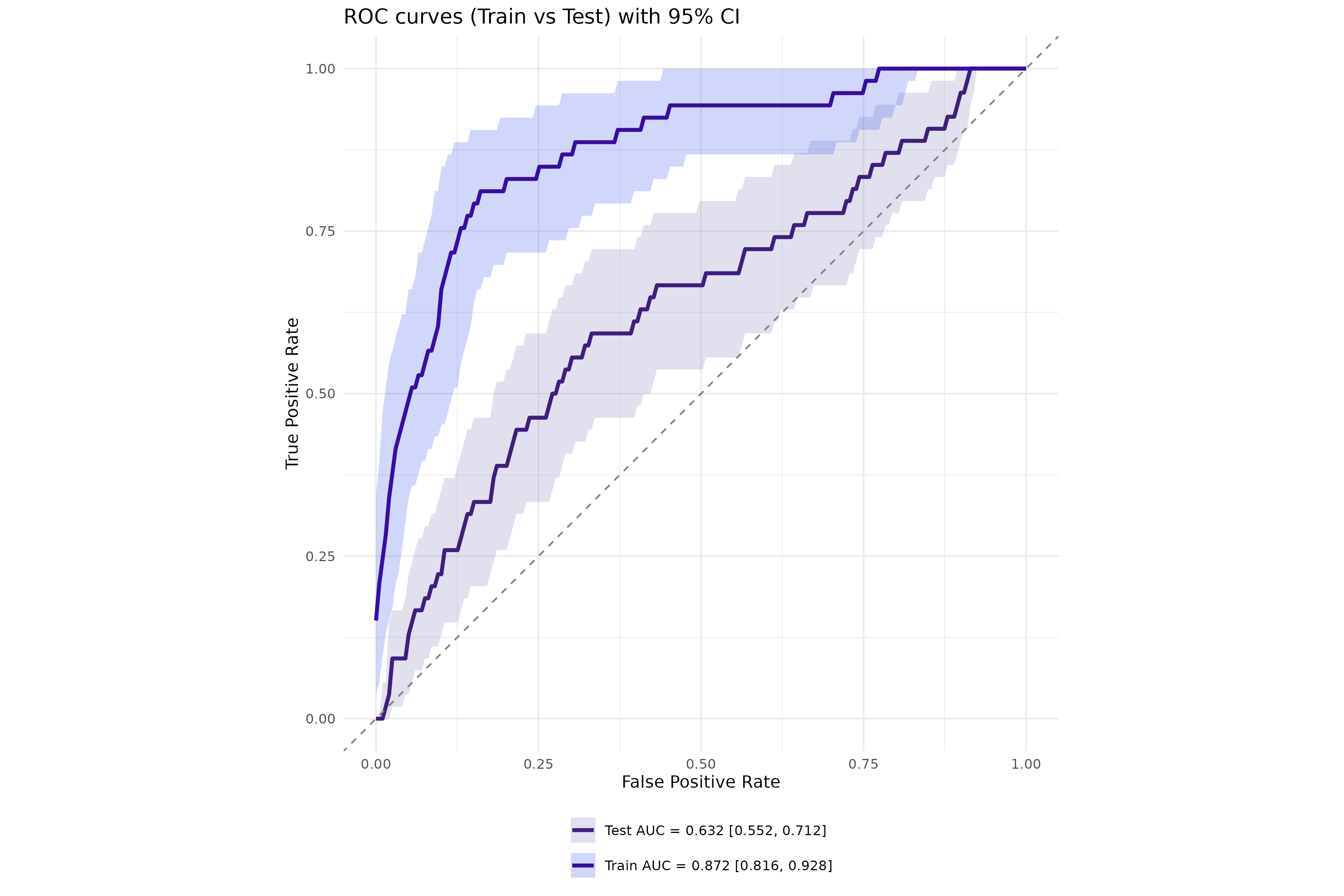}
\caption{ROC curves for the T2D risk prediction model based on 34
network-central proteins, evaluated on the balanced training set
($n=270$; 54 cases, 216 controls) and the population-representative
independent test cohort ($n=9{,}555$). Shaded bands indicate 95\%
pointwise confidence intervals for sensitivity at fixed specificity
values.}
\label{fig:S3}
\end{figure}

A standard logistic regression model, via \texttt{glm} package~\citep{venables2013modern}
link, was trained on the balanced training cohort ($n=270$; 54 incident
T2D cases and 216 controls matched 1:4 by sex and age) using the 34
network-central proteins identified through combined centrality analysis
as predictors. Model performance was evaluated on the independent test cohort
($n=9{,}555$; T2D incidence 0.5\%), producing a test AUC of 63.2\%
(95\% CI: 55.2--71.2\%) for predicting 5-year T2D onset
(Figure~\ref{fig:S3}). The high training AUC reflects the balanced
design of the training set (20\% cases vs 0.5\% in the population),
which is expected and not indicative of overfitting. However, the test AUC on
the population-representative cohort is the relevant measure of
generalisation. This performance should be interpreted in context: the
model uses only proteomic network-derived features without clinical risk
factors, and the 5-year prediction horizon is clinically demanding.
Incorporating network-central proteins into an augmented model alongside
standard clinical variables (as in QDiabetes-2018) represents a natural
next step.

\subsection*{Supplementary figures and Tables}

\begin{figure*}[htbp]
\centering
\includegraphics[width=0.75\textwidth]{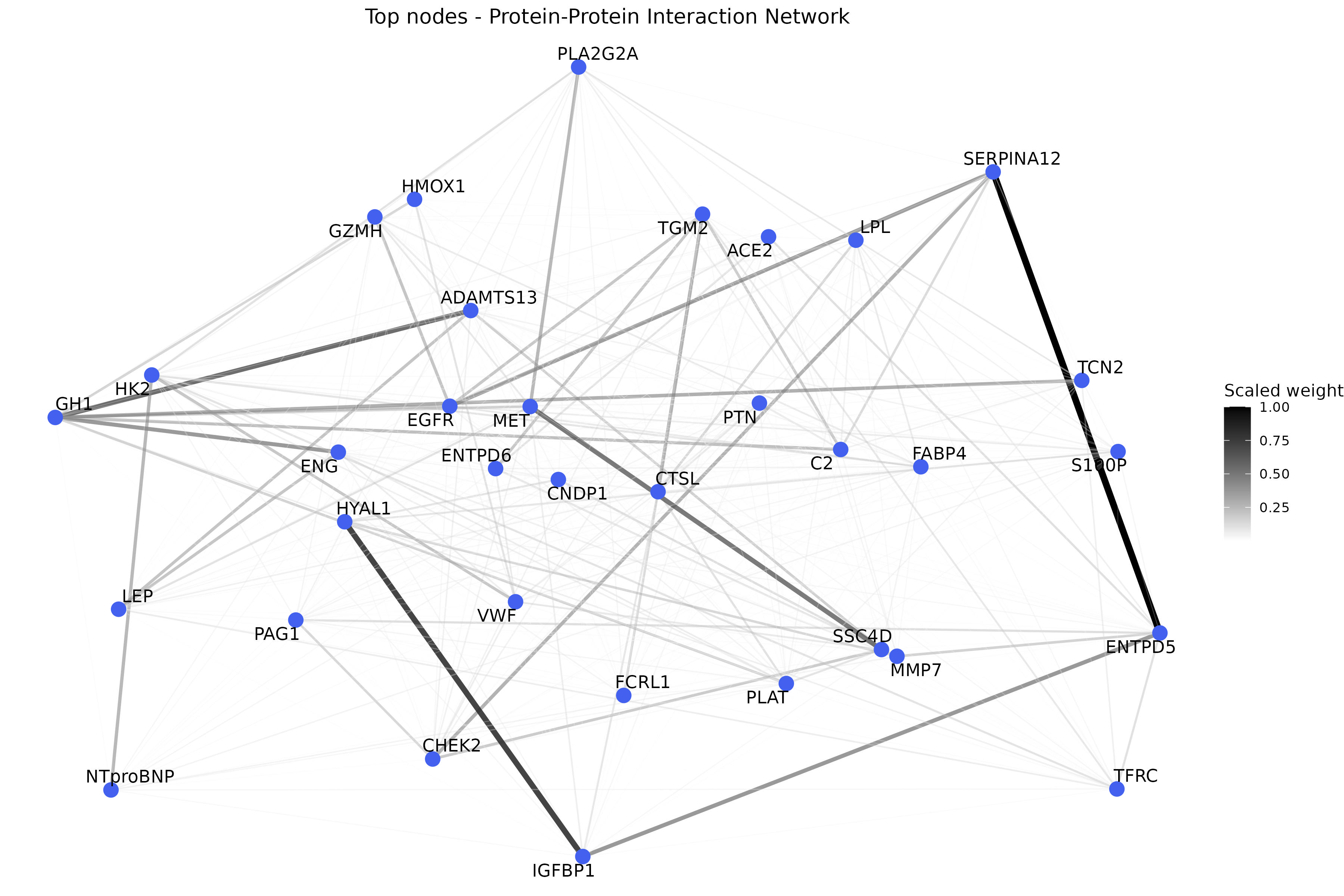}
\caption{Subnetwork of top-ranked central proteins.
Network visualisation focusing on the top 10\% highest-ranked proteins
based on combined centrality measures (degree, eigenvector, betweenness,
PageRank) intersected with the top 10\% hub-score proteins, and their
interconnections. Node labels show protein gene names. Edge colours
represent interaction strength (partial correlation). This subnetwork
represents the most influential proteins in the T2D-associated network
and primary candidates for biomarker development and therapeutic
targeting.}
\label{fig:S1}
\end{figure*}

\begin{figure*}[htbp]
\centering
\includegraphics[width=0.75\textwidth]{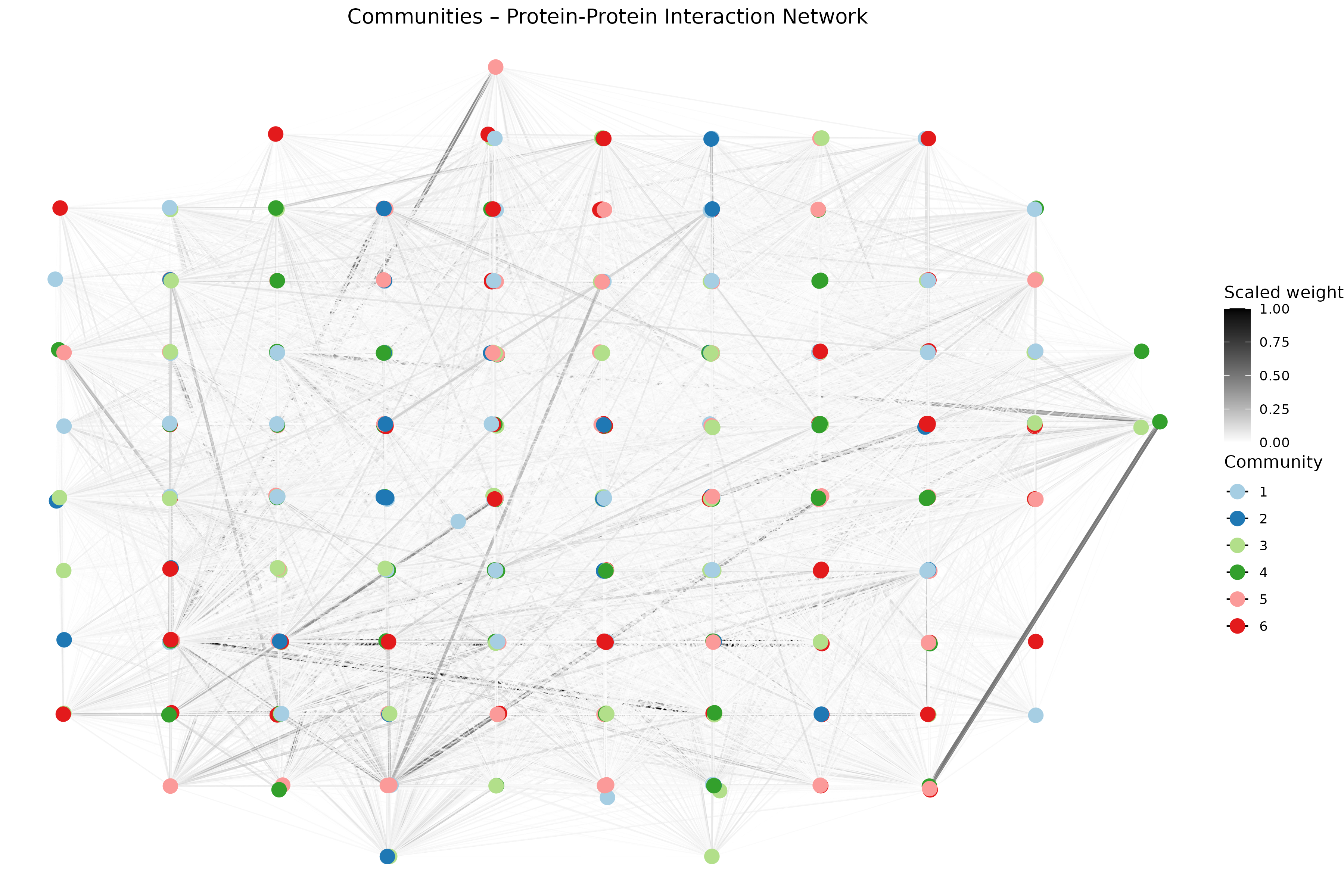}
\caption{Community structure of the T2D-associated protein
interaction network.
Network visualisation showing the six protein communities identified
through fast greedy modularity optimisation on the full T2D differential
network $\hat{\mathbf{A}}^{\mathrm{T2D}}$. Each node represents a
protein, with node colour indicating community membership (communities
1-6 shown in the legend). Edge colour reflects interaction strength.
Communities correspond to groups of proteins with particularly
strong mutual interactions within the T2D-perturbed network context.}
\label{fig:S2}
\end{figure*}


\begin{table*}[htbp]
\caption{Complete list of network-central proteins identified
through combined centrality analysis (Table~S1). Proteins are
identified by their gene symbol (Gene column). T2D association status and
gene-disease association scores are from the DisGeNET
database~\citep{pinero2016disgenet}: DPI (Disease Pleiotropy Index,
range 0-1; higher values indicate association with more diseases);
DSI (Disease Specificity Index, range 0-1; higher values indicate
more T2D-specific association); score (evidence-weighted association
strength, 0-1). Differential expression was tested between non-diabetic
individuals and incident T2D cases; fold change is reported for
significantly differentially expressed proteins only.
$\uparrow$: increased in T2D; $\downarrow$: decreased in T2D.
Proteins with no differential expression but high centrality (\textbf{bold})
represent network-based candidates not captured by standard differential approaches.}
\label{tab:S1}
\begin{tabular*}{\textwidth}{@{\extracolsep\fill}lcccccc@{\extracolsep\fill}}
\toprule
Gene & T2D in & DPI & DSI & DisGeNET & Diff.\ & Fold \\
     & DisGeNET &     &     & score    & expressed & change \\
\midrule
ACE2      & Yes & 0.91 & 0.40 & 0.90 & Yes & $1.15\uparrow$  \\
ADAMTS13  & Yes & 0.91 & 0.45 & 0.35 & Yes & $1.02\uparrow$  \\
C2        & Yes & 0.87 & 0.46 & 0.25 & Yes & $1.01\uparrow$  \\
\textbf{CHEK2}     & Yes & 0.87 & 0.41 & 0.10 & \textbf{No}  & --  \\
\textbf{CNDP1}     & Yes & 0.70 & 0.59 & 0.80 & \textbf{No}  & --  \\
CTSL      & Yes & 0.87 & 0.46 & 0.30 & Yes & $1.08\uparrow$  \\
EGFR      & Yes & 0.91 & 0.27 & 0.85 & Yes & $0.97\downarrow$  \\
\textbf{ENG}       & Yes & 0.91 & 0.39 & 0.45 & \textbf{No}  & --  \\
ENTPD5    & No  & --   & --   & --   & Yes & $0.98\downarrow$  \\
ENTPD6    & No  & --   & --   & --   & Yes & $1.01\uparrow$  \\
FABP4     & Yes & 0.87 & 0.44 & 0.55 & Yes & $1.14\uparrow$  \\
FCRL1     & No  & --   & --   & --   & Yes & $1.01\uparrow$  \\
GH1       & Yes & 0.96 & 0.32 & 1.00 & Yes & $0.86\downarrow$  \\
GZMH      & No  & --   & --   & --   & Yes & $1.04\uparrow$  \\
\textbf{HK2}       & Yes & 0.87 & 0.47 & 0.80 & \textbf{No}  & --  \\
HMOX1     & Yes & 0.91 & 0.36 & 0.95 & Yes & $1.00\uparrow$  \\
HYAL1     & Yes & 0.87 & 0.51 & 0.30 & Yes & $1.03\uparrow$  \\
IGFBP1    & Yes & 0.91 & 0.43 & 0.55 & Yes & $0.79\downarrow$  \\
LEP       & Yes & 0.96 & 0.32 & 1.00 & Yes & $1.15\uparrow$  \\
LPL       & Yes & 0.87 & 0.43 & 1.00 & Yes & $0.94\downarrow$  \\
MET       & Yes & 0.91 & 0.34 & 0.35 & Yes & $0.91\downarrow$  \\
MMP7      & Yes & 0.91 & 0.40 & 0.30 & Yes & $1.04\uparrow$  \\
\textbf{NTproBNP}  & No  & --   & --   & --   & \textbf{No}  & --  \\
PAG1      & Yes & 0.91 & 0.50 & 0.15 & Yes & $1.04\uparrow$  \\
PLA2G2A   & Yes & 0.91 & 0.48 & 0.20 & Yes & $1.03\uparrow$  \\
PLAT      & Yes & 0.96 & 0.38 & 0.90 & Yes & $1.15\uparrow$  \\
PTN       & No  & --   & --   & --   & Yes & $1.04\uparrow$  \\
S100P     & Yes & 0.87 & 0.49 & 0.20 & Yes & $1.04\uparrow$  \\
SERPINA12 & Yes & 0.87 & 0.53 & 0.45 & Yes & $1.06\uparrow$  \\
SSC4D     & No  & --   & --   & --   & Yes & $1.17\uparrow$  \\
TCN2      & Yes & 0.83 & 0.50 & 0.15 & Yes & $1.02\uparrow$  \\
TFRC      & Yes & 0.91 & 0.38 & 0.50 & Yes & $1.04\uparrow$  \\
TGM2      & Yes & 0.91 & 0.43 & 0.85 & Yes & $1.01\uparrow$  \\
VWF       & Yes & 0.96 & 0.35 & 0.60 & Yes & $1.03\uparrow$  \\
\multicolumn{7}{l}{\small 27/34 proteins (79\%) have documented T2D
  association in DisGeNET.}\\
\multicolumn{7}{l}{\small \textbf{Bold}: proteins with no differential
  expression (network-based candidates only).}\\
\end{tabular*}
\end{table*}

\begin{table*}[htbp]
\caption{Community membership and KEGG pathway enrichment summary. For each of the six communities identified by fast greedy
modularity optimisation, the table reports the number of member proteins,
the most central proteins in the community, and significantly enriched KEGG
pathway categories (adjusted $p \leq 0.05$). Community~1 had no
significant KEGG enrichments.}
\label{tab:S2}
\begin{tabular*}{\textwidth}{@{\extracolsep\fill}p{0.04\textwidth}
  p{0.06\textwidth}p{0.25\textwidth}p{0.55\textwidth}@{\extracolsep\fill}}
\toprule
Comm. & Size & Representative & Enriched KEGG categories \\
      & ($p$) & proteins & (adjusted $p \leq 0.05$) \\
\midrule
1 & 59 & TGM2, CTSL, ENTPD6, NOTCH1 & None \\
2 & 34 & S100P, TCN2, PCSK9, GP2 & Digestive system; Immune system; Carbohydrate metabolism; Immune disease \\
3 & 72 & HK2, NTproBNP, VWF, TFRC, F9, PTN, FCRL1  & Signaling molecules and interaction; Eukaryotic cellular community; Signal transduction; Immune system; Endocrine and metabolic disease; Infectious disease: viral; Energy metabolism; Transport and catabolism; 	
Infectious disease: parasitic; Digestive system \\
4 & 63 & IGFBP1, HYAL1, C2, LPL, TINAGL1, FETUB & Energy metabolism; Signalling molecules and interaction \\
5 & 72 & GH1, SSC4D, ADAMTS13, MET, EGFR, HMOX1, ENG, PLAT, GZMH, PLA2G2A, FABP4  & Signaling molecules and interaction; Cardiovascular disease; Cancer: overview; Cell growth and death; Signal transduction; Infectious disease: parasitic; Immune system; Digestive system; Infectious disease: viral; Lipid metabolism; Infectious disease: bacterial; 	
Cancer: specific types; Endocrine and metabolic disease; Endocrine system\\
6 & 66 & SERPINA12, ENTPD5, LEP, ACE2, MMP7, CNDP1, CHEK2, PAG1 & Signalling molecules and interaction; Transport and catabolism; Amino acid metabolism; Infectious disease: parasitic; Development and regeneration; Endocrine system \\
\end{tabular*}
\end{table*}